# Stability-guided reinforcement learning control of confined side-by-side cylinder wakes

Yongliang Yang[1], Yizhe Chen[1], Hui Tang[2], Guangrui Sun[3], Pengtao Wang[1], Rui Guo[1], Xuerao He[4,*]
[1]School of Mechanical Engineering, Nanjing University of Science and Technology, Nanjing 210094, China
[2]Department of Mechanical Engineering, The Hong Kong Polytechnic University, Kowloon, Hong Kong, China
[3]School of Energy Science and Engineering, Harbin Institute of Technology, Harbin 150001, China
[4]School of Mathematics, Monash University, VIC 3800, Australia
[a] Author to whom correspondence should be addressed: xuerao.he@monash.edu

**Abstract.** This study investigates the instability mechanisms and active-jet control of the wake of side-by-side cylinders in a channel, where the flow is simultaneously affected by wall confinement, gap-flow acceleration and wake–wake coupling. Through global stability analysis and direct numerical simulations, five distinct eigenmodes undergoing Hopf bifurcations are identified. Phase diagrams of these eigenmodes are established in the spacing-ratio--Reynolds-number (*L*/*d*-*Re*) and blockage-ratio--Reynolds-number ($\beta$-*Re*) parameter planes. And five corresponding nonlinear wake patterns arising from these linear instabilities are identified sensitivity analysis reveals the effects of the *Re*, *L*/*d* and $\beta$ on the spatial distribution of the 'wavemaker' and its relationship with flow stability. The physics-informed deep reinforcement learning based active flow control demonstrates that placing sensors in the 'wavemaker' region significantly improves the control efficiency. An enhanced reward function proposed in this study improves training convergence and robustness to the penalty parameter. To the best of our knowledge, the present study not only completely suppresses wake oscillations associated with Hopf bifurcation, but also provides the first demonstration of controlling bistability-induced flip-flopping and intermittent symmetry breaking, as well as chaotic wakes at high values of both $\beta$ and *Re*, driving these flows towards the unstable base state while achieving drag reduction. In particular, the restoration of flow symmetry represents an important advance in controlling complex multimode wake dynamics.

## 1. Introduction

Micro- to millimetre-scale channels are widely used in microreactors for fine chemical synthesis[1-4] because of their ease of integration, high productivity, product uniformity and precise controllability of reaction processes. In engineering applications, in addition to modular scale-up through the parallelization of multiple channels, another direct and effective way to increase reactor throughput is to increase the flow velocity within the channels. However, this inevitably increases the Reynolds number and may trigger flow instabilities, giving rise to undesirable behaviours such as symmetry breaking, flow oscillations and even chaotic states. Such instabilities are particularly detrimental to fine chemical synthesis, which requires precise flow control, as they may promote the local accumulation of process materials in narrow channels, potentially leading to clogging[4-7] and associated safety hazards. To stabilize unsteady flows in high-throughput micro- to millimetre-scale channels, the present work investigates active flow control (AFC) by combining flow-stability physical indicators with deep reinforcement learning (DRL) on a canonical microreactor configuration of two side-by-side cylinders, allowing systematic variation of wall confinement, gap flows, and wake interactions.

### 1.1 Flow instability mechanisms

In practical engineering applications, such as heat-exchanger tube bundles and subsea risers of offshore platforms, the flow past multiple circular cylinders is affected by several factors, including the Reynolds number, cylinder spacing, arrangement and angle of incidence[8]. For example, at intermediate spacings between the side-by-side cylinders, gap-flow acceleration and wake–wake coupling have comparable strengths and compete to dominate the flow field. This may lead to symmetry-breaking wake[9-10] (caused by pitchfork bifurcation) and periodically oscillating wake (caused by Hopf-bifurcation), as well as the flip-flopping behavior[11-13], in which the asymmetric deflection direction of the gap flow switches intermittently. However, the effect of wall confinement on side-by-side cylinder wakes in micro- to millimetre-scale channels remains unclear, even though such channel-flow configurations are common in industrial applications. In the absence of directly relevant studies, we draw on related work on uniform flow past two side-by-side cylinders to gain a partial understanding of the flow mechanisms involved.

Zdravkovich et al.[14] reviewed the hydrodynamic forces and vortex structures, and pointed out that both are considerably more complex than those for a single cylinder, being governed primarily by the cylinder arrangement and the Reynolds number. Sumner[15] showed that, as the spacing ratio of two side-by-side cylinders increases from 1 to 3, the flow successively undergoes three wake regimes: a single-bluff-body wake, an asymmetrical or biased gap flow and parallel vortex streets. The transition boundaries between these regimes were also identified, together with their connection to Kelvin–Helmholtz shear-layer instability. Where, for the biased gap flow, the two cylinders experience pronounced differences in force characteristics and dominant frequencies, while the direction of gap-flow bias switches randomly, a behaviour associated with shear-layer instability. In parallel vortex streets regime, the two vortex streets observed in the near wake merge into a single vortex street farther downstream. Carini et al.[11-13] demonstrated that the flip-flopping behavior of the wake behind side-by-side circular cylinders at low Reynolds numbers is not a bistable switching process induced by external random perturbations, but rather originates from an intrinsic secondary instability of the system, which further enriches the nonlinear wake dynamics. Ren et al.[16-17] systematically investigated the wake evolution of flow past side-by-side circular cylinders from two-dimensional to three-dimensional states. They mapped in detail the boundaries of six wake regimes in the gap-ratio–Reynolds-number parameter plane, including two previously unreported three-dimensional instability modes, and revealed the hysteresis associated with their subcritical transition nature. These findings highlight the remarkable complexity of side-by-side-cylinder wakes and the substantial challenge this poses for flow control. Does wall confinement, then, make side-by-side cylinder wakes more complex or enhance their stability? The answer has direct implications for the difficulty of flow control, making this one of the key questions addressed in this study.

Regarding the application of physical insights from sensitivity analysis to flow control, Giannetti et al.[18] first showed, through the overlap of direct and adjoint perturbation modes, that the wavemaker of the cylinder wake instability is in a small region within the wake. In other words, it is not located on the cylinder surface, where the adjoint mode attains its maximum, nor far downstream, where the direct mode is largest. Luchini et al.[19] reviewed how the combination of direct and adjoint modes can identify the location and strength of the wavemaker, a result with direct implications for flow control. Recently, Wei & Sun[20] showed how gap spacing and counter-rotation modify the instability-core distribution and neutral stability curves of side-by-side cylinders in an unbounded domain. Furthermore, in porous media flows[21], adjoint-based sensitivity analysis identifies the locations where placing a small cylinder produces the largest shift of the leading global mode. That is, sensitivity analysis provides a map for passive flow control, indicating where intervention is expected to be most efficient. However, the effectiveness of using this physical information to guide active flow control remains to be confirmed.

### 1.2 AFC with reinforcement learning

In recent years, synthetic-jet AFC has attracted considerable attention[22-24], as passive methods partly struggle with complex, dynamically evolving flows. AFC seeks to suppress vortex shedding, reduce drag or regulate flow symmetry through real-time actuation[25-28], but remains challenging because of the high-dimensional, strongly coupled nonlinear dynamics. DRL addresses this challenge by combining neural-network feature extraction with reinforcement-learning decision-making for high-dimensional flow states and sequential control[29-32]. For example, Rabault et al.[33] first applied DRL to a confined cylinder wake at $Re$=100, suppressing vortex shedding and reducing drag by 8% through two surface jets. Tang et al.[34-35] subsequently achieved approximately 30% drag reduction in weakly turbulent confined-cylinder flows. Using cylinder rotation, Ren et al.[36] reduced lock-in VIV amplitude by 99.6%. Fan et al.[37] achieved up to 30% drag reduction using two rotating auxiliary cylinders, and Sonoda et al.[38] obtained 37% skin-friction reduction in turbulent channel flow through wall blowing and suction. Font et al.[27] further showed that DRL can outperform periodic forcing by generating multifrequency actuation and can autonomously rediscover streamwise-vortex-based mechanisms for flow reattachment in high-$Re$ three-dimensional flows.

Recent advances in machine learning for fluid mechanics suggest embedding physical or mathematical constraints and prior knowledge rather than relying solely on brute-force search, thereby reducing the search space, guiding learning towards physically meaningful solutions and accelerating convergence[39-41]. For example, Li & Zhang et al. incorporated stability physical indicators into a DRL framework and, using only 10 probes, effectively suppressed wake oscillations behind a single cylinder in a confined channel[42], and while improved control performance for a more challenging convectively unstable flow[43]. For an unstable extensional flow with a saddle point, exploiting geometric symmetry improved training efficiency and mitigated inertia-induced response delays while successfully guiding a droplet to the centre of a four-roll mill[44]. Using Proper Orthogonal Decomposition physical information to optimize sensor distribution, Zhao et al.[45] developed a interpretable policy for suppressing lift fluctuations in tandem-cylinder wakes, leading to faster convergence and consistently improved performance. R. Ma et al.[46-47] identified the optimal jet-orifice location for suppressing periodic vortex shedding behind a circular cylinder using physical insights from resolvent analysis[48] and the Coandă effect, providing important guidance for control-strategy design. However, the generalizability of this finding to more complex flows involving multimode instabilities remains to be investigated. More recently, Lori et al.[49] proposed a physics-enhanced machine-learning framework and developed a spray-cooling prediction model that spans different fluids, geometries and flow conditions. The model achieved a prediction error below 2.3%, substantially outperforming empirical correlations, and was further used to inversely optimize the parameter space towards maximum heat-transfer performance.

The advances reviewed above demonstrate the potential of DRL for complex flow-control problems and provide useful insights for developing an efficient control strategy for the present configuration.

### 1.3 Contribution and organization

To the best of the authors' knowledge, the effects of confinement on the nonlinear wake dynamics and global stability of side-by-side circular cylinders in a channel remain insufficiently understood, and stability-analysis based DRL control for suppressing vortex shedding and reducing drag in such flows has not yet been reported. Therefore, the present study investigates how the Reynolds number and geometric parameters, including the blockage ratio and spacing ratio, affect the nonlinear dynamics, stability and structural sensitivity of the flow. The objective is to provide physics-based prior information for DRL-based AFC and to develop more effective control strategies.

This paper is organized as follows. Section 2 presents the problem formulation, the numerical method and the AFC framework. Section 3 is devoted to the global linear stability analysis (LSA) and sensitivity analysis. Section 4 analyses the control performance for confined flow past two side-by-side cylinders based on flow-stability indicators and DRL. Section 5 summarizes the main conclusions.

## 2. Problem formulation and method

This section first introduces the geometry and governing equations. It then explains how stability information guides sensor placement, actuator design and reward formulation for DRL-based AFC. Details of the numerical implementation, verification and validation are provided in Appendices A and B.

### 2.1 The nonlinear flow system

A schematic of two side-by-side cylinders in a channel flow is shown in Fig. 1. The line connecting the centres of the two cylinders is perpendicular to the direction of the incoming flow. A Cartesian coordinate system is adopted, with its origin located at the geometrical centre of symmetry of the cylinder pair. The $x$-axis is aligned with the flow direction, and the $y$-axis is defined according to the right-hand rule. The centre-to-centre distance between the two cylinders is $L$, and the diameter of each cylinder is $d$. The channel height is $H$ ($H/2$=1 is nondimensional), while the streamwise dimension of the computational domain is determined through a grid-independence study.

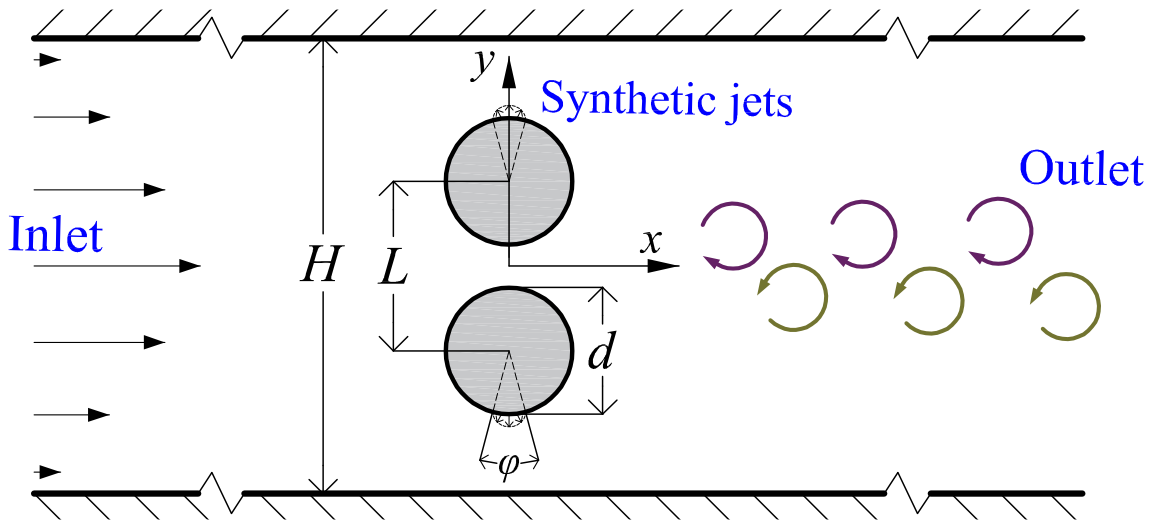

Fig. 1 Schematic of two side-by-side cylinders in a channel flow, not to scale.

The spacing ratio is $L/d$. A blockage ratio, $\beta$=2$d/H$, is defined to characterize the degree of confinement in the channel. The Reynolds number $Re$ and the Strouhal number $St$ are defined as

$$Re = \frac{U_{\max} d}{\nu}, \quad St = \frac{fd}{U_{\max}}, \tag{1}$$

where, $U_{\max}$ is the maximum inlet velocity, ν is the kinematic viscosity, and $f$ denotes the vortex-shedding frequency obtained from direct numerical simulations (DNS).

The drag $F_d$ and lift $F_l$ on a single cylinder are defined as the streamwise and transverse components, respectively, of the resultant hydrodynamic force. They are computed as

$$F_d = \int_C \left( \rho\nu \frac{\partial \boldsymbol{U}_\tau}{\partial \boldsymbol{n}} \boldsymbol{n}_y - P\boldsymbol{n}_x \right) \mathrm{d}C , \quad F_l = -\int_C \left( \rho\nu \frac{\partial \boldsymbol{U}_\tau}{\partial \boldsymbol{n}} \boldsymbol{n}_x + P\boldsymbol{n}_y \right) \mathrm{d}C , \tag{2}$$

where $C$ denotes the cylinder surface, ($\boldsymbol{n}_x$, $\boldsymbol{n}_y$) are the components of the outward unit normal vector in the streamwise and transverse directions, respectively, and $U_\tau$ denotes the tangential velocity component. The instantaneous drag coefficient $c_d$ and lift coefficient $c_l$ for a single cylinder are then defined as

$$c_d = \frac{F_d}{1/2 U_{\max}^2 d}, \quad c_l = \frac{F_l}{1/2 U_{\max}^2 d} . \tag{3}$$

The present work adopts two synthetic jets, a type of zero-net-mass-flux jets, for flow control. These two jets are located at the top of the upper cylinder and at the bottom of the lower cylinder, respectively, as shown in Fig. 1, and are oriented normal to the cylinder surfaces. The width of each synthetic jet is set to $\varphi=\pi/18$, and the jet velocity is prescribed using a cosine-like profile[33]:

$$\left(v_{jx}, v_{jy}\right)_{\text{upper}}^{\mathrm{T}} = Q_{\mathrm{u}}(t)\frac{2\pi}{\varphi d^2}\cos\left(\frac{\pi}{\varphi}\left(\alpha-\frac{1}{2}\pi\right)\right)(x,y)^{\mathrm{T}}, \quad \left(v_{jx}, v_{jy}\right)_{\text{lower}}^{\mathrm{T}} = Q_l(t)\frac{2\pi}{\varphi d^2}\cos\left(\frac{\pi}{\varphi}\left(\alpha-\frac{3}{2}\pi\right)\right)(x,y)^{\mathrm{T}}, \tag{4}$$

where, $v_{jx}$ and $v_{jy}$ are the velocity components in the $x$ and $y$ directions, respectively; $\alpha$ is the angular coordinate, in radians, of an arbitrary point $(x, y)$ on the jet boundary; and the flow rates of the upper and lower jets are controlled by adjusting the values of scalar quantities $Q_{\mathrm{u}}$ and $Q_l$, respectively.

The nonlinear flow system is governed by the two-dimensional viscous incompressible Navier–Stokes equations:

$$\frac{\partial \boldsymbol{U}}{\partial t} + \boldsymbol{U}\cdot\nabla\boldsymbol{U} = -\nabla P + \frac{\beta}{Re}\nabla^2\boldsymbol{U}, \qquad \nabla\cdot\boldsymbol{U} = 0 , \tag{5}$$

where, $\boldsymbol{U}$ denotes the velocity and $P$ the pressure. A velocity-in boundary condition is imposed at the channel inlet, where the velocity profile is prescribed by a parabolic distribution $U_x$=1-$y^2$ and $U_y$=0. Here, ($U_x$, $U_y$) are the streamwise and transverse velocity components at the inlet boundary. No-slip boundary conditions are applied on the cylinder surfaces and the channel walls. At the outlet, a zero-gradient outflow boundary condition $\left(-P\boldsymbol{I} + \frac{\beta}{Re}\nabla\boldsymbol{U}\right)\cdot\boldsymbol{n} = 0$ is imposed to avoid non-physical backflow or reflected waves, where, $\boldsymbol{I}$ is the identity tensor and $\boldsymbol{n}$ is the unit vector outward normal to the boundary face. The numerical implementation and verification are provided in Appendix A.

## 2.2 The linear flow system

### 2.2.1 Linearized Navier–Stokes equations

To analyse the linear instability mechanism in the wake of side-by-side cylinders in a confined domain, the incompressible Navier–Stokes equations (5) are linearized. Following our previous work[50-52, 21], the asymptotic-expansion approach of flow field is decomposed as

$$(\boldsymbol{U}, P) = (\boldsymbol{U}_b, P_b) + \varepsilon(\boldsymbol{u}, p) + O(\varepsilon^2) , \tag{6}$$

where infinitesimal $\varepsilon \ll 1$. $(\boldsymbol{u}, p)$ denote the perturbation quantities. $(\boldsymbol{U}_b, P_b)$ denote the time-independent base flow. In the present confined flow past side-by-side cylinders, the flow undergoes a Hopf bifurcation and evolves from a steady state to a periodic oscillatory state. Therefore, for cases with $Re<Re_c$ ($Re_c$ is critical Reynolds number), base flow can be obtained directly from DNS. For cases with ($Re\geq Re_c$), the base flow is obtained using the selective frequency damping (SFD) method[50-51].

Substituting Eq. (6) into Eq. (5), simplifying the resulting equations and retaining only the first-order terms yields the following linearized equations:

$$\frac{\partial \boldsymbol{u}}{\partial t} + \nabla\boldsymbol{u}\cdot\boldsymbol{U}_b + \nabla\boldsymbol{U}_b\cdot\boldsymbol{u} = -\nabla p + \frac{\beta}{Re}\nabla^2\boldsymbol{u}, \qquad \nabla\cdot\boldsymbol{u} = 0 . \tag{7}$$

The boundaries for the linear direct problem (7) are the same as those in the nonlinear Navier–Stokes equations (5), except that the inlet boundary condition for the velocity is a Dirichlet type with $u$=0. The general solution of Eq. (7) can then be expressed as

$$(\boldsymbol{u}(x,y,t),\, p(x,y,t))^{\mathrm{T}} = (\hat{\boldsymbol{u}}(x,y),\, \hat{p}(x,y))^{\mathrm{T}} \mathrm{e}^{\sigma t}, \tag{8}$$

where, $\sigma = \lambda + \mathrm{i}\omega$ is the complex eigenvalue, whose real part $\lambda$ represents the temporal growth rate of the perturbation and imaginary part $\omega$ gives the eigenfrequency. The eigenmode $\hat{q} = (\hat{\boldsymbol{u}}, \hat{p})^{\mathrm{T}}$ describes the spatial distribution of the perturbation amplitude and phase at a given position ($x$, $y$).

Substitution of (8) into the linearized equations (9) then gives

$$\boldsymbol{B}\hat{\boldsymbol{q}} = \sigma\hat{\boldsymbol{q}}, \qquad \boldsymbol{B} = \begin{pmatrix} -\boldsymbol{U}_b \cdot \nabla - \nabla \boldsymbol{U}_b \cdot + \beta Re^{-1}\nabla^2 & -\nabla \\ \nabla \cdot & 0 \end{pmatrix}, \tag{9}$$

where, $\boldsymbol{B}$ is the Jacobian matrix that depends on the base state $(\boldsymbol{U}_b, P_b)$. The global linear stability problem is therefore transformed into a generalized eigenvalue problem, which is solved in the present study using a matrix-free method, namely the implicitly restarted Arnoldi method (IRAM). Rather than explicitly construct the Jacobian matrix $\boldsymbol{B}$, IRAM generates a Krylov subspace and constructs an orthonormal basis using Gram–Schmidt orthogonalization, thereby obtaining a much smaller Hessenberg matrix that approximates the Jacobian matrix. The stability of the fluid system can then be determined from the leading eigenvalue of the Hessenberg matrix. Specifically, the flow is unstable when $\lambda > 0$, and the perturbation amplitude grows with time; the flow is stable when $\lambda < 0$, and the perturbation amplitude decays with time; and the flow is marginally stable when $\lambda = 0$.

### 2.2.2 Structural sensitivity analysis

Based on the adjoint theory of the linearized Navier–Stokes equations, structural sensitivity analysis is performed, with particular emphasis on the characteristics of the wavemaker region and its variation with the governing parameters. Following the studies of Giannetti & Luchini[18] and Marquet et al.[53], the adjoint modes associated with the velocity and pressure perturbations ($\hat{\boldsymbol{u}}$, $\hat{p}$), denoted by $\hat{\boldsymbol{u}}^+$ and $\hat{p}^+$, are introduced to formulate the adjoint form of the linearized Navier–Stokes equations as follows

$$\sigma^*\hat{\boldsymbol{u}}^+ - \nabla\hat{\boldsymbol{u}}^+ \cdot \boldsymbol{U}_b + (\nabla \boldsymbol{U}_b)^T \cdot \hat{\boldsymbol{u}}^+ = -\nabla\hat{p}^+ + \frac{\beta}{Re}\nabla^2\hat{\boldsymbol{u}}^+, \qquad \nabla \cdot \hat{\boldsymbol{u}}^+ = 0. \tag{10}$$

Where, $\sigma^*$ is the eigenvalue of the adjoint mode. For the adjoint system, homogeneous boundary conditions $\hat{\boldsymbol{u}}^+ = 0$ are imposed at the inlet and on the walls, while the corresponding adjoint outflow condition $\hat{p}^+\boldsymbol{n} - \beta Re^{-1}\left(\nabla\hat{\boldsymbol{u}}^+\right)\cdot\boldsymbol{n} = \left(\boldsymbol{U}_b \cdot \boldsymbol{n}\right)\hat{\boldsymbol{u}}^+$ is applied at the outlet. The generalized eigenfunctions of the adjoint modes are also solved using the IRAM.

The direct and adjoint modes exhibit substantial differences in their spatial structures, and examining either eigenfunction alone is therefore insufficient to identify the core mechanism of the instability. Following Giannetti & Luchini[18], the overlap between the eigenvector of the direct mode $\hat{\boldsymbol{u}}$ and that of the adjoint mode $\hat{\boldsymbol{u}}^+$, is used to determine the spatial distribution where the coupling between the direct and adjoint velocity is strongest. This distribution identifies the location and strength of the most sensitive region of the flow, namely the wavemaker region $\boldsymbol{\eta} = |\hat{\boldsymbol{u}}||\hat{\boldsymbol{u}}^+| / \langle \hat{\boldsymbol{u}}, \hat{\boldsymbol{u}}^+ \rangle$. The IRAM and structural sensitivity analysis codes developed in our previous work[50-51] was adapted for the present study, and its validation is provided in Appendix B.

### 2.3 DRL-based AFC

For the confined side-by-side cylinder flows considered here, nonlinear dynamics, complex wake interactions and intermittent wake switching in bistable regimes pose challenges for passive and conventional open-loop control strategies. Benefiting from the mechanism of environment interaction and reward feedback, DRL can autonomously learn effective control policies for different flow states[30, 40]. Meanwhile, by exploiting the powerful nonlinear representation capability of neural networks, DRL effectively alleviates the curse of dimensionality associated with high-dimensional state spaces[54-55]. Therefore, DRL provides a promising approach for addressing the high-dimensional, nonlinear and dynamic control problem considered in the present study.

#### 2.3.1 DRL-based AFC framework

The present DRL-based AFC framework is illustrated in Fig. 2. Through real-time closed-loop interaction between the agent and the environment, the control policy (action) is iteratively optimized based on the feedback reward $r_t$ after executing the control action, aiming to maximize the long-term cumulative discounted reward and thereby suppress flow instability and oscillations. The agent employs separate multilayer perceptrons (MLP) for the actor and critic. The actor consists of two 512-neuron hidden layers and parameterizes the policy distribution over continuous control actions, whereas the critic consists of two 32-neuron hidden layers and estimates the state-value function used for advantage estimation in PPO. At time $t$, the agent receives the state $s_t$, consisting of velocity measurements ($U_x$, $U_y$) from the sensors (green dots in Fig. 2), and feeds it into the actor MLP. The network parameterizes a probability distribution over the synthetic-jet flow-rate actions, from which the control action $a_t$ is sampled, thereby enabling state-dependent adaptive control.

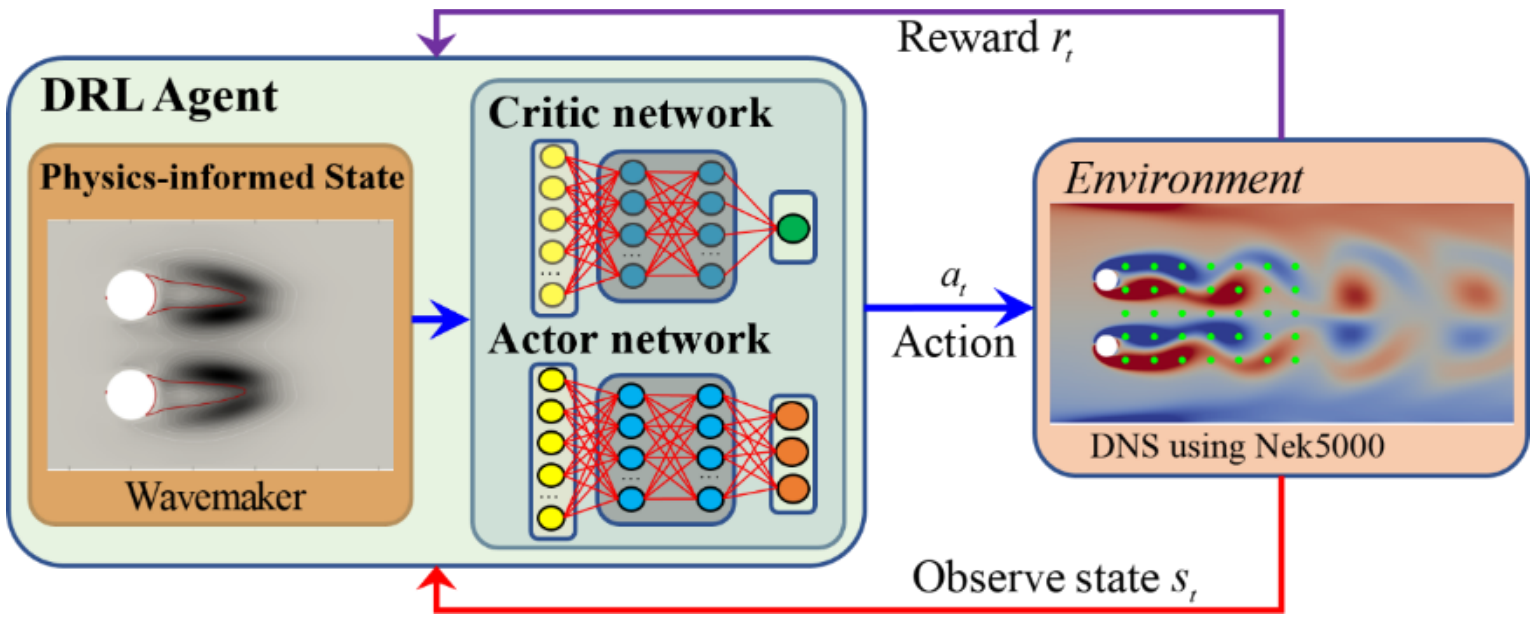

Fig. 2 DRL-based AFC framework

After executing $a_t$, the agent receives the updated state $s_{t+1}$ and reward $r_t$ from the environment. Based on the new state $s_{t+1}$, the agent estimates the state-value function $V(s_{t+1})$ of the next action, and approximately evaluates the expected return at time $t$, thereby assessing the quality of the current action $a_t$. The agent then updates the policy network by combining the current reward with the future return, providing guidance for subsequent iterations.

To prevent excessive changes in the jet, which may introduce non-physical abrupt variations in the flow velocity and pressure fields and cause DNS divergence, thereby interrupting the DRL training process, the jet-action update strategy is adopted as follows[33] $Q_{\text{jet}}^{(t+1)} = Q_{\text{jet}}^{(t)} + \gamma \times (Q_{\text{action}} - Q_{\text{jet}}^{(t)})$, where $Q_{\text{jet}}^{(t)}$ denotes the jet flow rate at the previous time step, $Q_{\text{jet}}^{(t+1)}$ represents the updated jet flow rate at the current time step. The smoothing coefficient is set to $\gamma$=0.1. $Q_{\text{action}}$ is the action set by the proximal policy optimization (PPO) agent for the current 100 DNS time steps. The control action is constrained by the condition $Q_{\text{u}} + Q_l$ =0, ensuring that no additional mass is introduced into the flow field. Therefore, only the flow rate of the upper jet is adjusted, while the lower jet maintains the same magnitude with an opposite direction.

The update interval of the control action is identical to that used for dynamic mode decomposition (DMD[56], an equation-free, data-driven method, can be used for nonlinear flow stability analysis.) calculation and is denoted as $\Delta t_a$. To ensure that the control strategy can effectively respond to changes in the system state while reducing disturbances caused by frequent actuation updates, following the study of Li & Zhang[42], $\Delta t_a$ is selected as 6–8% of the vortex-shedding period in present work. We set 80 actions in each episode, thus composed of 8 physical time units (i.e. 4-6 vortex shedding periods). A fixed number of 500 episodes (i.e. 4000 time units) is used during the training stage.

Within the actor–critic framework, the PPO algorithm proposed by Schulman et al.[57] is adopted. PPO provides stable policy updates, is model-free, supports parallel training and continuous action spaces, and exhibits good generalization capability, making it suitable for active control of the complex high-dimensional flow system considered in present work. The Adam optimizer is employed for neural-network training with a fixed learning rate of 0.001.

#### 2.3.2 Design of the reward function

The nonlinear saturated flow state after the Hopf bifurcation obtained from DNS is used as the initial state for active control. This initial state captures the essential dynamical characteristics of natural vortex shedding. Li & Zhang[42] demonstrated that active control based on machine-learning methods can drive the oscillatory nonlinear saturated flow field towards the unstable SFD base flow. This finding provides a physical basis for constructing the reward function and enables a more accurate quantitative evaluation of the control performance.

The basic reward function $R_w$ adopted in the present work is defined to minimize the oscillation energy in the flow field, as expressed in Eq. (11). It consists of three key components:

1) Reducing the kinetic-energy difference between the controlled flow field and the steady SFD base flow.

2) The maximum DMD growth rate $g$ is used in the present work to guide the control policy towards more stable flow states. This DMD-based indicator has been shown to be more effective than that obtained from the linear equations[42] and to exhibit good robustness in online drag-reduction control of cylinder wakes[58].

3) Introducing a penalty term $P_e$ to avoid slow learning caused by excessively small reward values, or gradient vanishing/explosion caused by excessively large rewards. The parameter $P_e$ is determined empirically, and its influence on the control performance is discussed in the following section.

$$R_w = P_e - E_S \mathrm{e}^g , \qquad E_S = \sum_{j=1}^{n_{\mathrm{node}}} \left( \left( U_x^j - U_{bx}^j \right)^2 + \left( U_y^j - U_{by}^j \right)^2 \right) . \tag{11}$$

Where $n_{\mathrm{node}}$ is the total number of grid nodes; ($U_{bx}$, $U_{by}$) is the steady SFD base flow; and $E_S$ is the sum of the fluctuation of kinetic energy in the whole computational domain. The variable $g$ denotes the maximum DMD growth rate, and is embedded into the reward function as physical information from the flow environment to improve the agent's learning efficiency in capturing the dynamic evolution of the flow field. $\mathrm{e}^g<1.0$ when $g<0$, corresponding to a stable flow, indicating that the associated control strategy is effective, and the reward increases. $\mathrm{e}^g=1.0$ when $g=0$, corresponding to a neutrally stable flow, and the reward set to the original kinetic-energy value. $\mathrm{e}^g>1.0$ when $g>0$, corresponding to an unstable flow, indicating that the associated control strategy is ineffective, the reward decreases.

Furthermore, to avoid excessive positive or negative reward values caused by the empirically determined parameter $P_e$, which may lead to over or insufficient exploration during training, a dynamic normalization strategy based on historical rewards is introduced into the reward-feedback design of the DRL framework. This approach aims to improve the convergence performance and efficiency of the DRL training process. The resulting enhanced reward $R_{ws}$ is defined as follows

$$R_{ws} = \frac{R_w^{np} - \mu_s}{\sigma_s}, \quad \sigma_s = \sqrt{\frac{1}{m}\sum_{k=np-m+1}^{np}\left(R_w^k - \mu_s\right)^2}, \quad \mu_s = \frac{1}{m}\sum_{k=np-m+1}^{np} R_w^k, \quad , \tag{12}$$

where $R_w^k$ represents the reward corresponding to the $k$-th action, $np$ denotes the number of control actions that have been executed, and $\mu_s$ is the mean value of the historical rewards. $\sigma_s$ is the standard deviation of the reward. The normalized reward $R_{ws}$ is obtained by centring the current reward $R_w^{np}$ with respect to the historical mean and scaling it by the standard deviation, thereby avoiding excessive concentration or fluctuation of reward values. In the present work, the mean and standard deviation are calculated from the historical rewards of the most recent $m$=80 actions and are updated online at each training step to adaptively adjust the reward value.

### 2.3.3 Validation of the proposed DRL-based AFC strategy

A supercritical case ($\beta$=0.25, $Re$=150) of flow past a single cylinder in confined domain is selected to validate the present flow-control framework and numerical code, and to provide a preliminary assessment of the control efficiency. The numerical configuration is identical to that used in Ref[42]. A total of 86 sensors are employed and distributed in the downstream region of the cylinder, as shown in the figure 19(b) of Ref[42].

Fig. 3 (a) shows the control results for the vortex-shedding kinetic energy under the base reward and enhanced reward functions with different values of the penalty parameter $P_e$. Here, ($P_e$=$P_1$=800), which is obtained by rounding up the difference between the instantaneous kinetic energy of the flow field at the initial control time ($t$=0) and the kinetic energy of the corresponding SFD base flow.

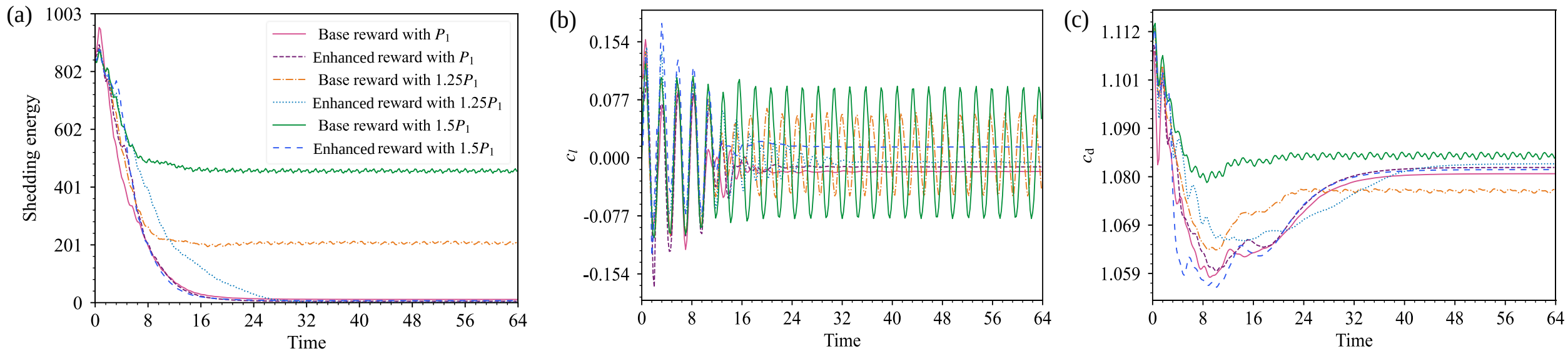

Fig. 3 Effect of $P_e$ on the control performance for confined flow past a single cylinder at $\beta$=0.25 and $Re$=150. Time histories of (a) the vortex shedding energy, (b) lift coefficients and (c) drag coefficients.

For the base reward function, after approximately 100 control actions have been executed (i.e., $t$>20), the vortex-shedding kinetic energy decreases and converges to its minimum value. For penalty parameters $P_e$=$P_1$, 1.25$P_1$ and 1.5$P_1$, the time-averaged vortex-shedding kinetic energy converges to approximately 0, 200 and 460, respectively. This indicates that, for $P_e$=1.25$P_1$ and 1.5$P_1$, the DRL agent does not learn an effective control policy, and periodic oscillations persist in the flow field, as shown in Fig. 4 ($d$, $f$). As can be seen from Eq. (11), the reward value increases linearly with the $P_e$. However, an excessively large $P_e$ causes the rewards fed back from all actions to remain positive, with only small variance. When processing such reward data, the agent cannot distinguish between favourable and unfavourable actions, becomes trapped in a local optimum and ceases exploration. Consequently, an optimal control policy cannot be obtained through training.

For the enhanced reward function, the penalty parameter is likewise set to $P_1$, $1.25P_1$ and $1.5P_1$. In all three cases, the flow oscillations can be suppressed. Moreover, the enhanced reward achieves a more complete suppression of the vortex-shedding kinetic energy than the basic reward. The minimum vortex-shedding kinetic energy obtained with the enhanced reward at $P_e=P_1$, $1.25P_1$ and $1.5P_1$ is reduced to 41%, 16% and 67%, respectively, of that obtained with the basic reward function at $P_e=P_1$. For $P_e=1.25P_1$, approximately 150 control actions are required for the vortex-shedding kinetic energy to reach its minimum value (i.e., $t$>30); however, this case gives the strongest suppression of the vortex-shedding kinetic energy.

The time histories of the lift coefficient in the testing stage are shown in Fig. 3 (*b*). When the basic reward function is used with $P_e=1.25P_1$ and $1.5P_1$, the amplitude of the lift coefficient is reduced, but the oscillations are not fully suppressed. In contrast, when the enhanced reward function is used, the lift oscillations are suppressed for all three penalty values. Among them, the case with $P_e=1.25P_1$ yields a lift coefficient closest to zero, with a value of -0.005.

The time histories of the drag coefficient in the testing stage are shown in Fig. 3 (c). In all cases, the drag coefficient first decreases and then increases, and the final drag remains lower than that of the uncontrolled flow, indicating that suppression of vortex shedding is beneficial for drag reduction. For the enhanced reward function with the three penalty values, the drag oscillations are all effectively suppressed. It is worth noting that, although the basic reward function with $P_e=1.25P_1$ does not completely suppress the oscillations, it achieves the best drag-reduction performance, with the time-averaged drag coefficient reduced by 4% relative to the uncontrolled case. This suggests that the relationship between vortex-shedding suppression and drag reduction is not strictly monotonic.

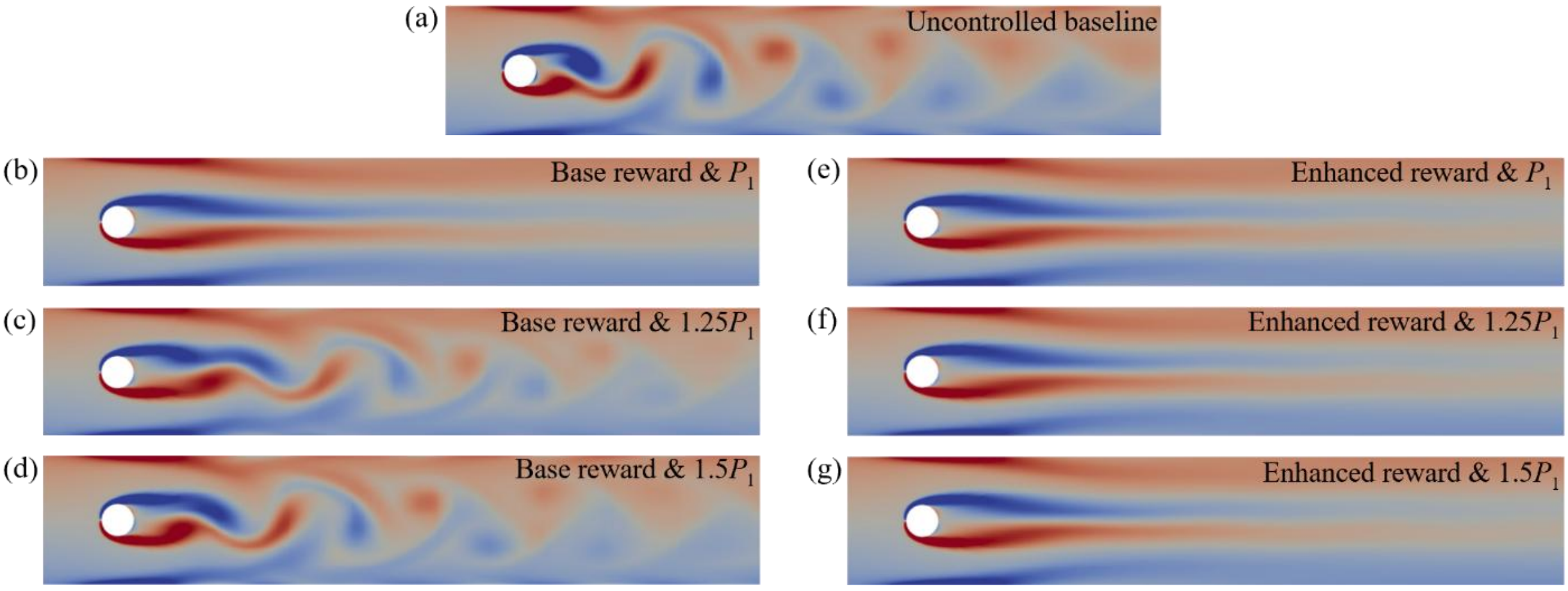

Fig. 4 Vorticity snapshots for confined flow past a single cylinder at $\beta$=0.25 and $Re$=150, with vorticity range: [-5,5]. The uncontrolled baseline (a) at $t$=0, and the controlled flows (b–g) are shown at $t$=64. The displayed region is $x$=[-2,10] and $y$=[-1,1].

The vorticity contours for the baseline (Fig. 3 *a*) and the controlled cases (Fig. 3 *b*–*g*) for the unsteady case ($\beta$=0.25, $Re$=150) are shown in Fig. 4. Where Fig. 4 (b) shows the vorticity field obtained using the basic reward function with $P_e=P_1$ at the controlled time $t$=64, where the flow oscillations are completely suppressed. Although oscillations are not entirely eliminated in the cases with the basic reward function at $P_e=1.25P_1$ and $P_e=1.5P_1$, as shown in panels (c) and (d), respectively, the oscillatory region behind the cylinder is smaller than that of the baseline. Panels (*e*), (*f*) and (*g*) show that, when the enhanced reward function is used, vortex shedding can be suppressed towards the SFD base flow for all three penalty values.

In summary, for the supercritical case ($\beta$=0.25, $Re$=150), approximately 1.83 times the critical value $Re_c$=82

reported by Li & Zhang[42], the nonlinear oscillations of the single-cylinder wake flow in a confined domain are successfully suppressed. This validates the effectiveness of the present control framework. The proposed enhanced reward function is less dependent on empirical tuning of the penalty parameter, substantially reducing the time cost associated with manual trial and error, while also achieving stronger suppression of flow oscillations. These results provide the basis for the subsequent flow-control study of side-by-side cylinders in a confined domain.

## 3. Flow stability and sensitivity analysis

This section considers two sets of side-by-side cylinder configurations: one with a fixed blockage ratio of $\beta$=0.2 and varying spacing ratio $L/d$, and the other with a fixed spacing ratio of $L/d$=2.0 and varying blockage ratio $\beta$. For each set, we examine how the varying parameter affects the neutral stability curves of the different modes and the spatial distribution of their wavemaker regions.

### 3.1 Effect of spacing ratio

At a fixed blockage ratio of $\beta$=0.2, the confined flow past side-by-side cylinders exhibits different flow structures as the spacing ratio increases. These flow structures have a direct influence on the flow stability. The effect of the spacing ratio on the neutral stability curves of the three modes is shown in Fig. 5, i.e., phase diagram of the dominant unstable modes. The grey region above the neutral curves denotes the linearly unstable regime after the Hopf bifurcation, where the nonlinear saturated state of the flow takes the form of unsteady vortex-shedding oscillations. The region below the neutral curves corresponds to the linearly stable regime, in which the wake remains steady. A comparison with the critical $Re$ curves reported by Carini et al.[12-13] for unconfined flow past side-by-side cylinders is also shown in Fig. 5(a).

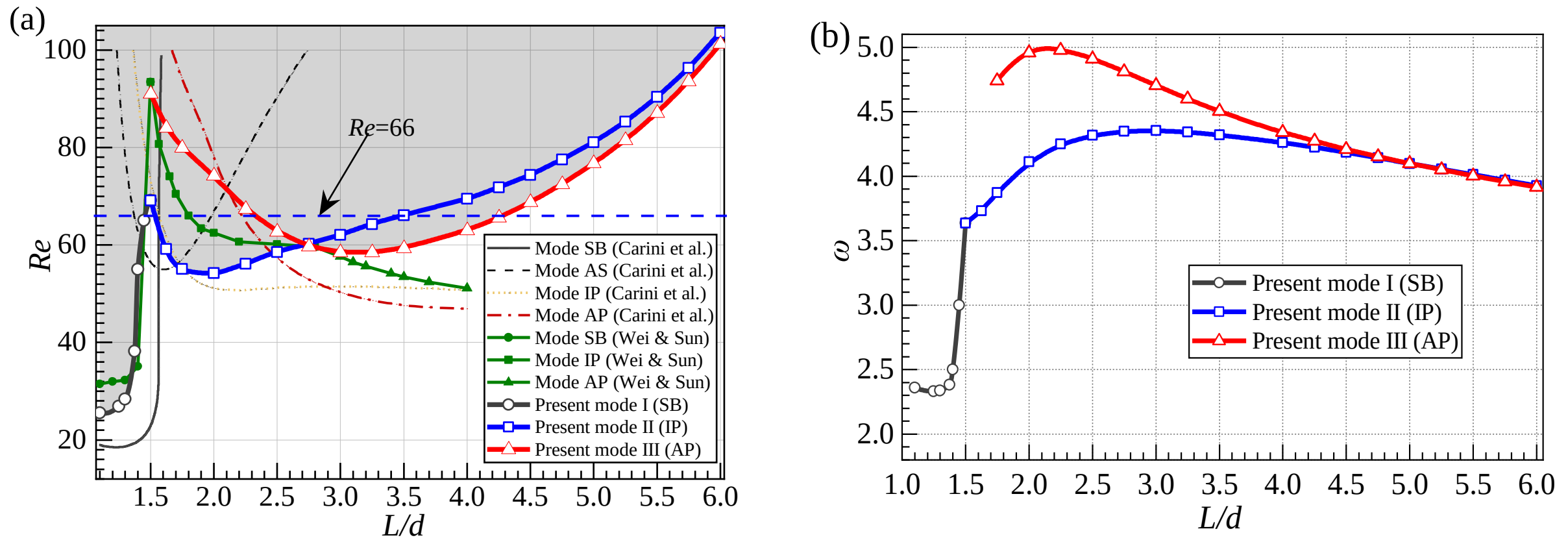

Fig. 5 The effect of spacing ratio $L/d$ on neutral stable curves in the $L/d$-$Re$ (panel a) and $L/d$-$\omega$ (panel b) parameter planes at $\beta$=0.2. The grey shaded area in panel (a) denotes unsteady flow patterns undergoing Hopf bifurcation. Compared with modes SB, AS, IP and AP for unconfined flow past side-by-side cylinders (Carini et al.[12-13]), and the modes SB, IP and AP for the flow past side-by-side counter-rotating cylinders at $\alpha$=0.5 (Wei & Sun[20]).

For $L/d$≤1.5, the low-frequency eigenmode I (SB) is the dominant mode. The $Re_c$ and the eigenfrequency of eigenmode I (SB) increase sharply as the spacing ratio is increased, which similar to the $L/d$-$Re_c$ relationship of mode SB reported by Carini et al.[12-13] for unconfined flow past side-by-side cylinders. This indicates that wall confinement has little effect on the flow stability under these conditions.

In the range (1.5<$L/d$<6.0), however, the behaviour differs from that in the unbounded-domain studies of Carini et al.[12-13]. As the two side-by-side cylinders move progressively closer to the channel walls, the $Re_c$ of eigenmodes II and III both decrease rapidly at first and then increase, indicating the stabilizing effect of the channel walls at large spacing ratios. Coincidentally, the codimension-two point[51]. of the modes IP and AP for flow past side-by-side counter-rotating cylinders[20] at $\alpha$=0.5 almost coincides with that of the present confined side-by-side cylinder flow, at approximately $Re$≈60, $L/d$=2.75. Therefore, active control through the above cylinder counter-rotation, and passive control through present variation of the $L/d$, may be regarded as two alternative routes to achieving the same control objective. When $L/d$>2.75, the growth rate of the present eignmode III exceeds that of eigenmode II, and eigenmode III replaces eigenmode II as the dominant mode.

Fig. 6 shows the drag–lift phase diagrams for five types of nonlinear saturated states at different $L/d$ and $Re$. For the cases ($L/d$=1.25, $Re$=66) and ($L/d$=4.0, $Re$=66), the $c_d$–$c_l$ phase diagrams of the upper and lower cylinders are symmetric with respect to each other (panel $b$). In the latter case, the time histories of $c_l$ are also symmetric about $c_l$=0 (panel $c$), so that the total lift coefficient, Tot. $c_l$, remains zero throughout (panel $a$). By contrast, for ($L/d$=1.25, $Re$=66), the lift-coefficient oscillations of the two cylinders are nearly synchronized, resulting in a symmetric '8'-shaped Tot. $c_d$–Tot. $c_l$ phase diagram. For the other three cases, the $c_d$–$c_l$ phase diagrams of the upper and lower cylinders are asymmetric and differ substantially in shape, leading to asymmetric Tot. $c_d$–Tot. $c_l$ phase diagrams. In addition, the phase diagram for ($L/d$=3.0, $Re$=66) suggests that the wake oscillation contains multiple frequency components.

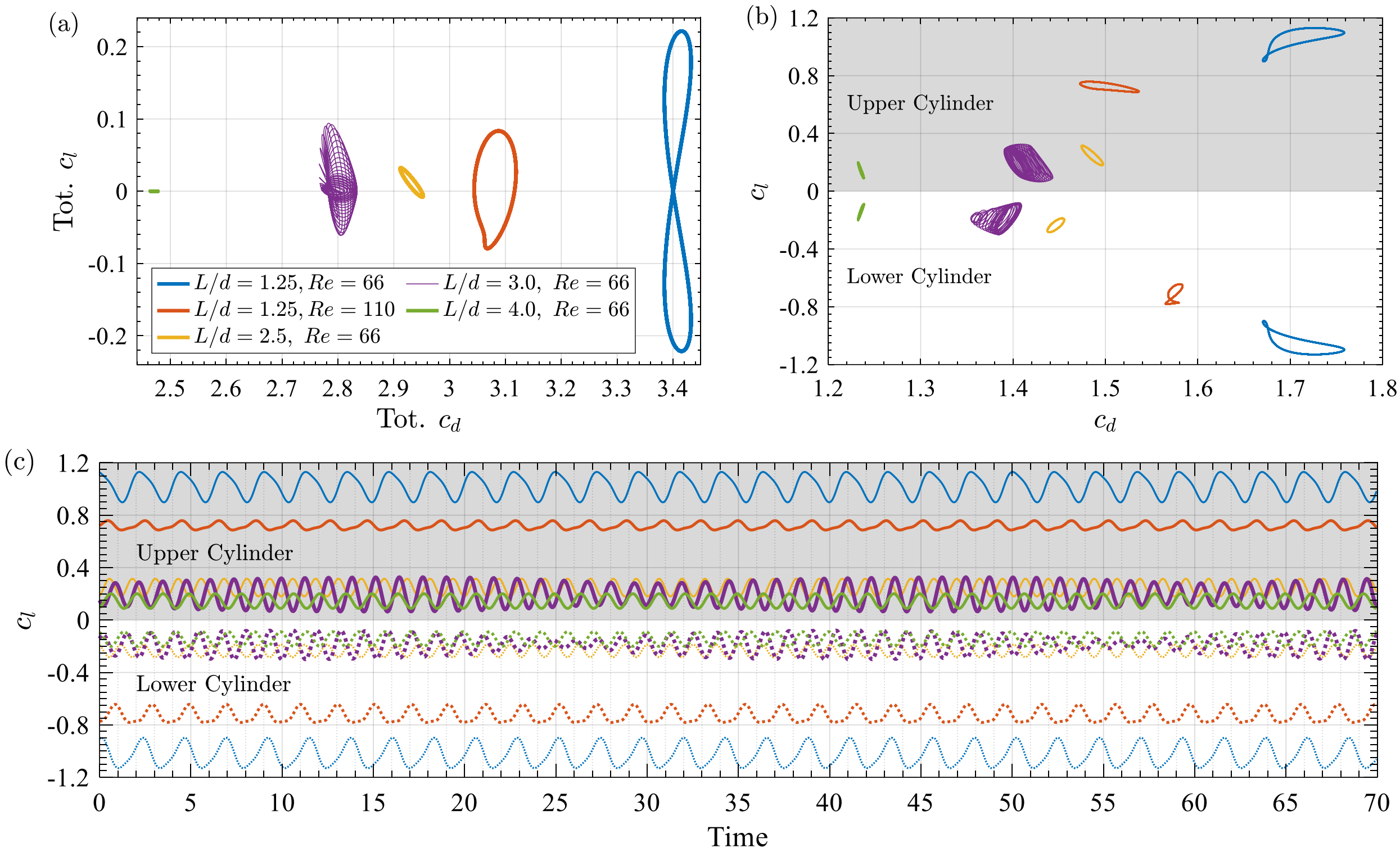

Fig. 6 The drag-lift coefficients of nonlinear saturated state at different spacing ratios $L/d$ and Reynolds numbers $Re$. Where 'Tot.' denotes the sum for the two cylinders. The grey-shaded and white regions correspond to the coefficient curves of the upper and lower cylinders, respectively.

The above neutral-stability and nonlinear saturated drag-lift phase diagrams reveal that $L/d$ is a primary control parameter governing transitions between flow modes for side-by-side cylinders in a channel. Based on these diagrams, four representative flow-control cases are selected in Section 4; all four lie on the blue dashed line in Fig. 5, which corresponds to $Re$=66.

### 3.1.1 Influence on global stability modes

**(A) Flow regime for $L/d$<1.5**

When the spacing ratio is small, eg. $L/d$<1.5, the gap flow is weak and the two cylinders behave like a single cylinder. The structures of eigenmodes 0 and I obtained from LSA are shown in Fig. 7 (a) and (b), respectively. These modes resemble the confined single-cylinder mode shown in Fig. B20, with no evident local features associated with the individual cylinders. Their symmetry is the same as that of the base flow; that is, symmetric about the centreline ($y$=0). Among them, eigenmode 0 remains linearly stable throughout the parameter space considered in the present study, and therefore does not give rise to nonlinear oscillations in the wake. As $L/d$ increases, the strengthened gap flow interacts with the shear layers shed from the outer sides of the side-by-side cylinders. This interaction weakens the vortices and leads to a rapid increase in $Re_c$ of the eigenmode I, as shown in Fig. 5 (a).

From DNS, the wake pattern is characterized by a single row of large-scale Kármán vortices shed alternately from the cylinder pair; this regime is referred to as wake pattern P-I in the present work. A representative instantaneous wake pattern for the case ($L/d$=1.25, $Re$=66) is shown in Fig. 7(c). The vorticity of the corresponding mean flow is symmetric about the centreline $y$=0. It's lift coefficients of the two cylinders oscillate in phase with nearly the same frequency and amplitude, while remaining on opposite sides of the zero-lift axis, corresponding to synchronized periodic oscillations with opposite-signed lift forces, as shown in Fig. 6.

Besides, at relatively high $Re$ (>~110), the flow becomes asymmetric in the gap region, called wake pattern P-I-A (see Fig. 7 d) in the present work. It is similar to the wake structure observed for side-by-side cylinders in an unbounded domain[13], where the single-bluff-body vortex shedding (termed mode SB in Ref.[13], undergoing a Hopf bifurcation) and the asymmetric steady flow (termed mode AS in Ref.[13], undergoing a pitchfork bifurcation) coexist. This asymmetric behaviour results in pronounced differences between the lift-coefficient waveforms of the upper and lower cylinders (see Fig. 6).

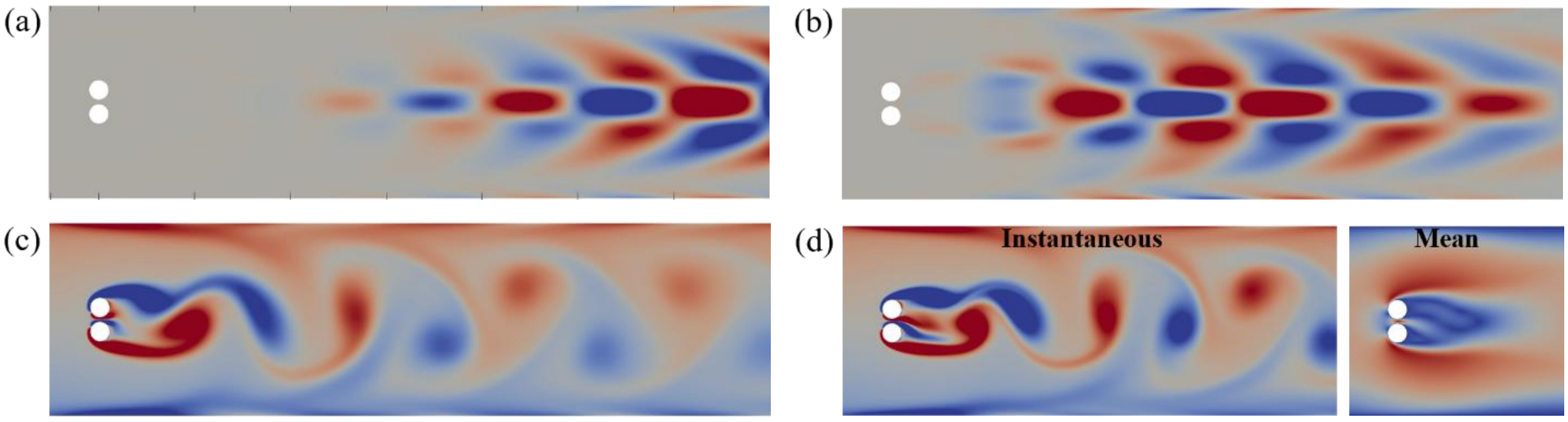

Fig. 7 Linear (a, b) and nonlinear (c, d) flow fields at a small spacing ratio of $L/d$=1.25. (a) Eigenmode 0 at $Re$=66 ($\sigma$=-0.24760, $\omega$=2.75545); (b) eigenmode I at $Re$=66 ($\sigma$=0.30358, $\omega$=2.36445), (c) wake pattern P-I at $Re$=66, and (d) instantaneous vorticity and time-averaged velocity fields of wake pattern P-I-A at $Re$=110. Vorticity range in panels (c, d): [-5,5].

**(B) Flow regime for 1.5<$L/d$≤2.75**

When 1.5<$L/d$≤2.75, the eigenmodes and the vorticity field of the nonlinear instantaneous flow for a representative case ($L/d$=2, $Re$=66) are shown in Fig. 8. LSA shows that eigenmode II, corresponding to the in-phase (IP) mode, first becomes unstable via a Hopf bifurcation, followed by eigenmode III, corresponding to the anti-phase (AP) mode. The vorticity field of eigenmode II (IP) is symmetric about the axis $y$=0, whereas that of eigenmode III (AP) is antisymmetric about $y$=0. In addition, these eigenmodes can be readily distinguished according to their eigenfrequencies: eigenmode I is a low-frequency mode, while eigenmodes II and III are high-frequency modes.

In the near wake, each cylinder generates its own vortex street, and the shedding from the two cylinders is in phase, indicating that the instability associated with each local cylinder begins to play a dominant role in the flow destabilization. In the intermediate and far wake, the two vortex streets intertwine and undergo vorticity annihilation, and the combined wake system is subsequently reorganized into a single row of larger size Kármán vortex street. The vorticity field of the corresponding mean flow remains symmetric about the axis $y$=0. As shown in Fig. 5(a), as the spacing increases, the gap flow is strengthened and the $Re_c$ initially decreases, reaching a minimum at $L/d$=2. With a further increase in $L/d$, the parallel channel walls begin to suppress the oscillatory shear layers on the outer sides of the cylinders, thereby stabilizing the flow and causing the $Re_c$ to increase again.

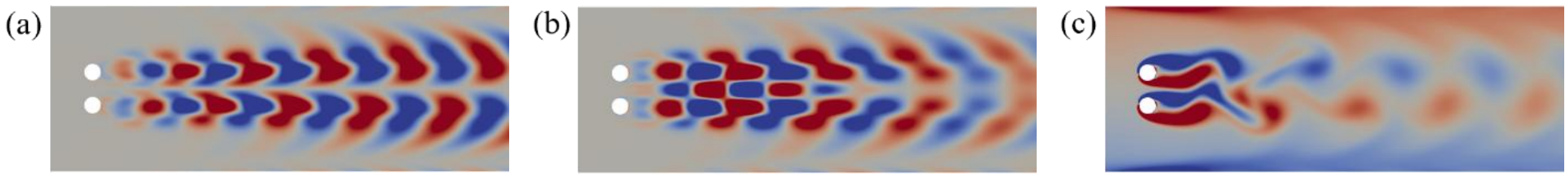

Fig. 8 Vorticity contours for case $L/d$=2.0 and $Re$=66, displayed over the streamwise range -0.5<$x$<5. (a) Eigenmode II (IP) ($\sigma$=0.243162, $\omega$=4.187099) and (b) eigenmode III (AP) ($\sigma$=0.113473, $\omega$=4.920604). (c) The instantaneous field of wake pattern P-II, where vorticity range: [-5,5].

Besides, when $L/d$=2.5, the gap flow becomes biased towards one side, producing a wide wake with strong oscillations and a narrow wake with weak oscillations. This regime, denoted as pattern P-III, is also observed at $L/d$=3.0, as shown in Fig. 9 (c). This asymmetric phenomenon, referred to as bistability in unbounded flow past two side-by-side cylinders[11], can give rise to flip-flopping behaviour.

**(C) Flow regime for *L/d*>2.75**

When $L/d$>2.75, the representative eigenmodes and nonlinear wake patterns are shown in Fig. 9. It is worth noting that, when eigenmode III (AP), becomes unstable first via a Hopf bifurcation, the resulting wake is characterized by two rows of vortices shed with a phase difference of 180°. As $L/d$ increase, the flow in the gap region gains more momentum, and the wake no longer forms a single vortex street. Instead, two rows of smaller-scale vortices develop behind the two cylinders, forming a vorticity field that is antisymmetric about axis $y$=0, which is referred to as the antisymmetric wake pattern P-III in the present work. Moreover, for the case $L/d$=3.0 and $Re$=66, intermittent switching occurs between the wake pattern P-III and the wake pattern P-IV.

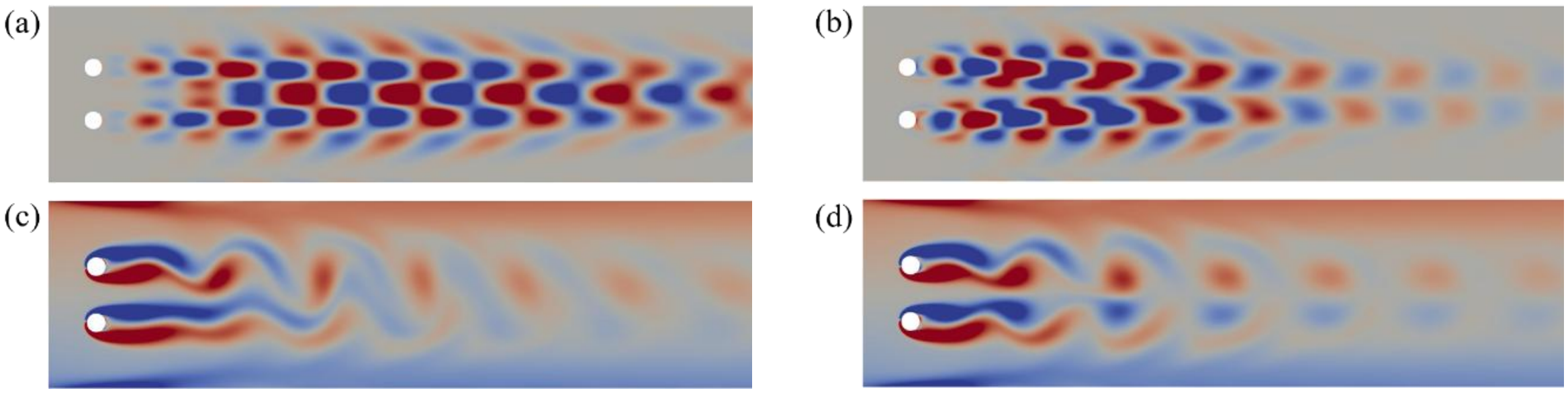

Fig. 9 Vorticity contours for $L/d$=3.0 and $Re$=66, displayed over the streamwise range -0.5<$x$<7. (a) Eigenmode II (IP) ($\sigma$=0.144382, $\omega$=4.733781) and (b) eigenmode III (AP) ($\sigma$=7.148166×$10^{-2}$, $\omega$=4.37314). The instantaneous field of (c) pattern P-III and (d) pattern P-IV, where the vorticity range: [-5, 5].

As $L/d$ is further increased, the intermittent switching between wake patterns P-III and P-IV disappears. For example, in the case with $L/d$=4.0 and $Re$=66, only wake pattern P-IV is observed, whereas the asymmetric wake pattern P-III is not found. LSA shows that, in this case, the antisymmetric eigenmode III ($\sigma$=5.27188×$10^{-2}$, $\omega$=4.35447) is unstable, while the symmetric eigenmode II ($\sigma$=-5.69024×$10^{-2}$, $\omega$=4.24391) remains stable. This suggests that intermittent switching is associated with the coexistence of competing unstable eigenmodes, and disappears when only one eigenmode remains unstable.

### 3.1.2 Influence on structural sensitivity

Boujo et al.[59] used sensitivity analysis to identify the optimal direction for modifying the recirculation length, and then determined whether the recirculation region should be shortened or elongated according to the Reynolds number, thereby enabling efficient active control of complex separated flows. Similarly, in this section, structural sensitivity analysis is performed at a fixed blockage ratio of $\beta$=0.2 to examine the effects of the spacing ratio and Reynolds number on the spatial distribution of the region most sensitive to perturbations, namely the wavemaker. This provides physical guidance for the design of efficient active-control strategies in subsection 2.3. Specifically, placing sensors in this region to collect flow-state information can influence the control policy in a more targeted and effective manner[42].

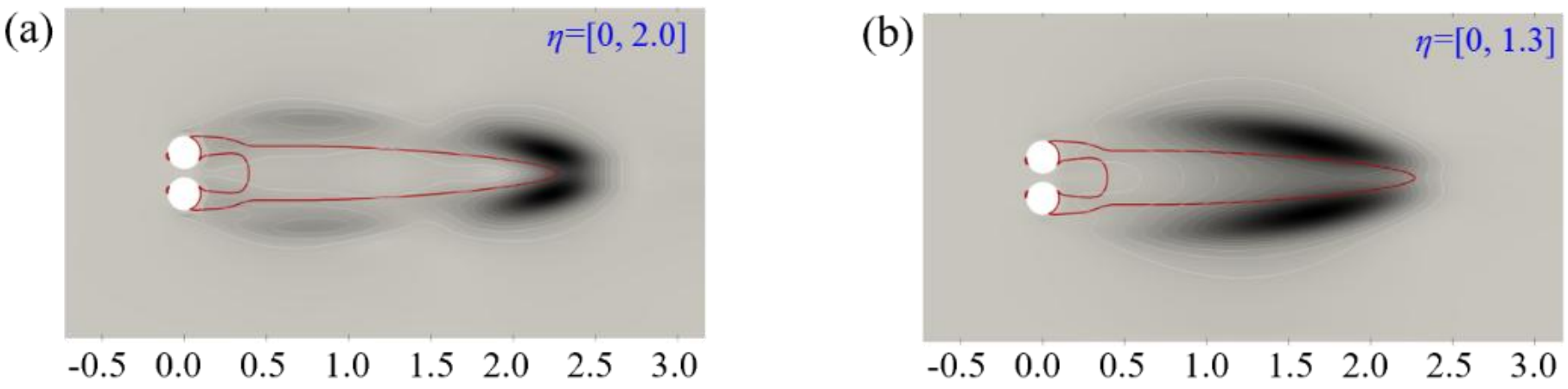

Fig. 10 Wavemaker distributions for (a) eigenmode 0 and (b) eigenmode I at $L/d$=1.25 and $Re$=66. Darker grey-scale regions indicate higher perturbation sensitivity. The region enclosed by the red solid line, $U_x(x, y)$=0, and the cylinder surfaces denotes the separation bubble of the base flow.

At a fixed $\beta$=0.2, Fig. 10, Tab. 1 and Tab. 2 show the spatial distributions of the wavemaker for the three eigenmodes at $1.25 \le L/d \le 5.0$ and $50 < Re < 120$, covering both subcritical and supercritical cases. It should be noted that, because $\eta$ is a scalar quantity, the wavemaker of the confined flow past side-by-side cylinders is symmetric about axis $y$=0, even when the vorticity of eigenmode III is antisymmetric.

The wavemaker of the present confined cylinder flow is located downstream of the cylinders, close to the recirculation region of the base flow, and the two regions overlap over a large area. Since the instability core is consistently anchored in the wake region, the upstream region of the cylinders plays a relatively minor role. So, priority is not given to monitoring the upstream region when suppressing vortex shedding in the confined wake.

As the gap spacing increases, the instability core gradually evolves from a single-body-like wavemaker, as observed for eigenmode I in Fig. 10 (b), into two weakly coupled localized structures, as shown by the wavemakers of eigenmodes II and III in Tab. 1. This marked distinction is favourable for independent control of the different instability modes. For example, if stronger suppression of the oscillations associated with eigenmode I is desired, more sensors can be placed farther downstream in the cylinder wake. By contrast, the wavemaker of eigenmodes II and III overlap strongly in both location and size. Therefore, the oscillation information associated with both eigenmodes II and III can be captured simultaneously using the same sensor-placement strategy.

For $\beta$=0.2 and $Re$=66, as shown in Tab. 1, the streamwise lengths of both the wavemaker and the base-flow recirculation region decrease as the $L/d$ increases. Combined with Fig. 5 (a), which shows that $Re_c$ increases with $L/d$ for $L/d \gtrsim 2.75$, these results suggest that shortening the streamwise length of wavemaker and the base flow recirculation regions is associated with enhanced flow stability.

It is worth noting that the maximum sensitivity (denoted by $\eta_{max}$) of eigenmode II is initially smaller than that of eigenmode III. However, as $L/d$ is gradually increased beyond 2.75, $\eta_{max}$ of eigenmode II exceeds that of eigenmode III, and the location of $\eta_{max}$ in the wavemaker region gradually shifts from the outer sides of the two cylinders to the inner gap region. Examination of the velocity field shows that this behaviour is related to the relative acceleration of the fluid in the gap and on the outer sides of the two cylinders. When the *L/d* is small, the flow in the gap region is hindered, while the fluid on the outer sides of the two cylinders experiences stronger acceleration. As *L/d* increases, the location of the maximum velocity gradually moves downstream towards the gap region, leading to a larger velocity gradient between the downstream gap flow and the recirculation region, and hence to a higher sensitivity to perturbations.

Tab. 1 Effect of *L/d* on the structural sensitivity at *Re*=66.

| Modes II (Sym) | Mode III (AS) |
| --- | --- |
| 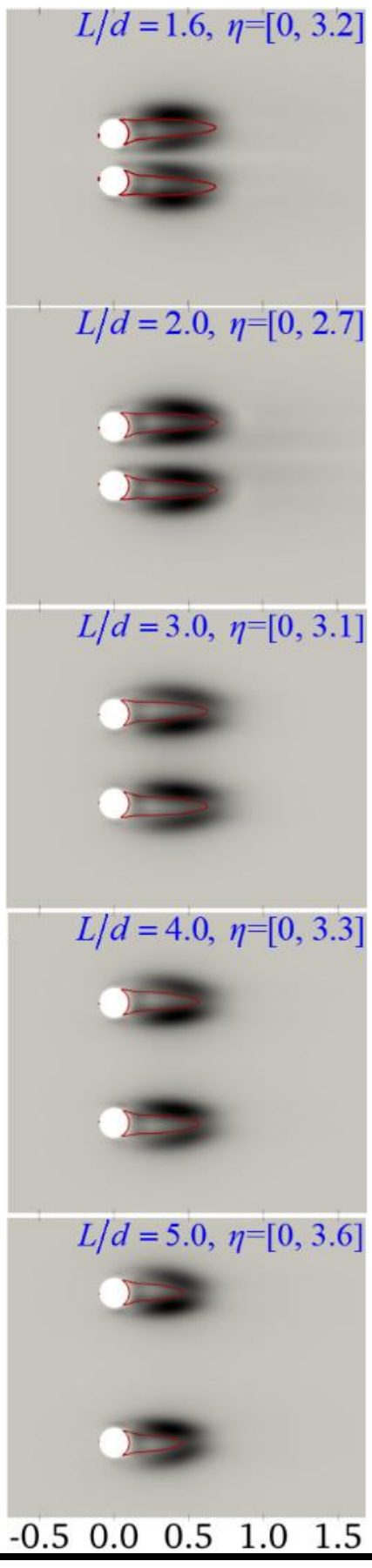  | 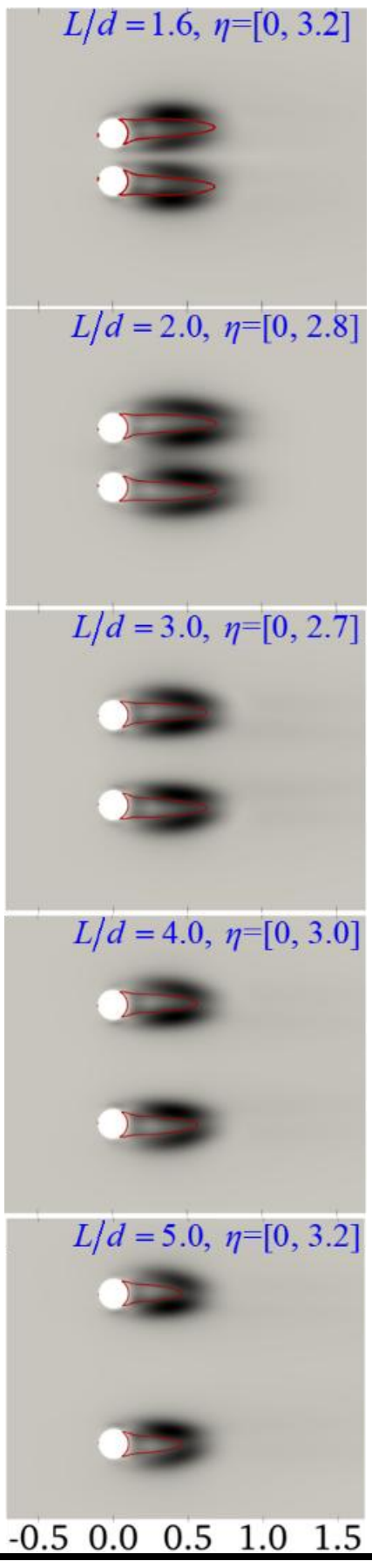  |

Tab. 2 Effect of *Re* on the structural sensitivity at *L/d*=4.

| Mode II (Sym) | Mode III (AS) |
| --- | --- |
| 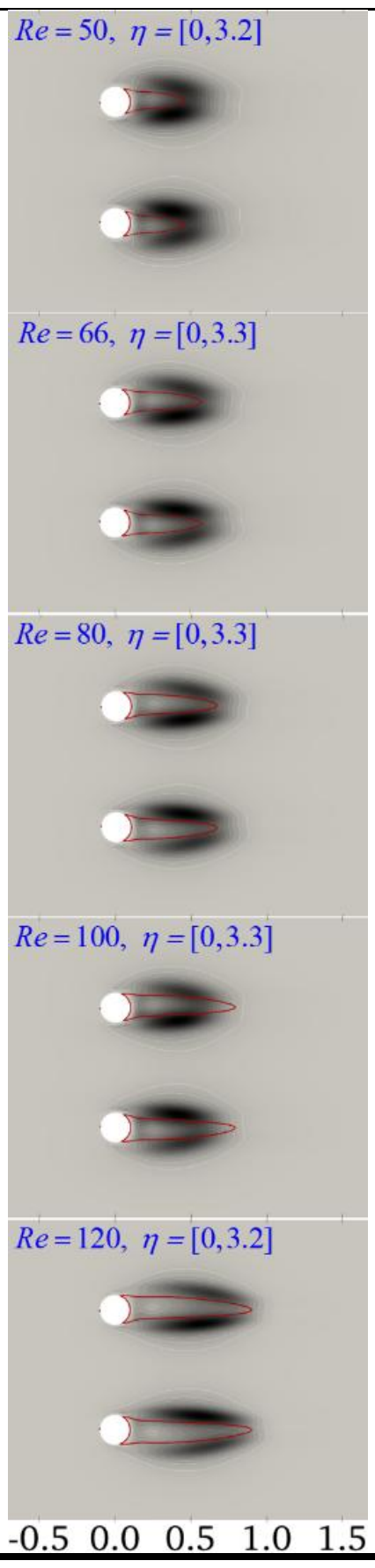  | 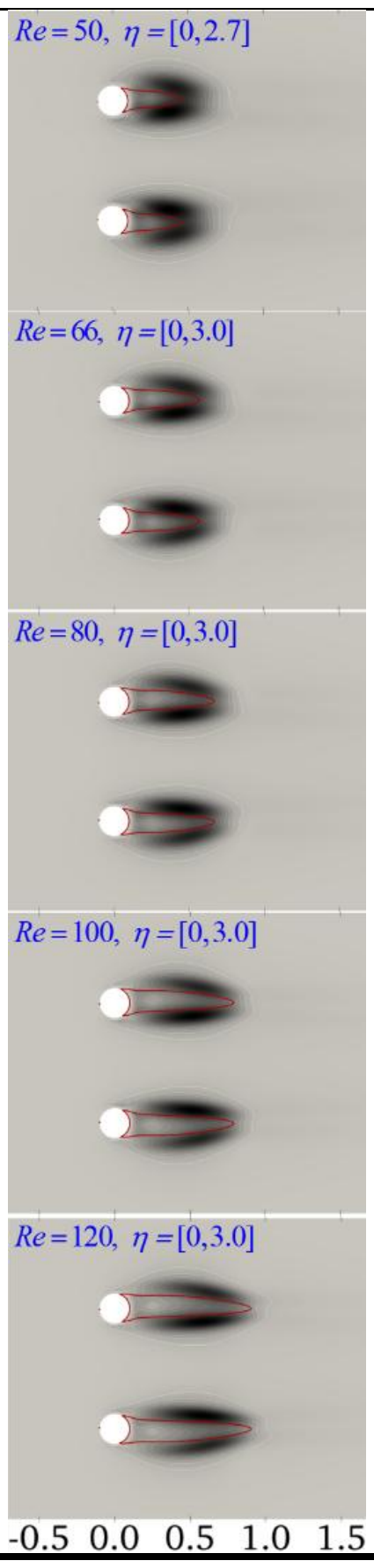  |

As shown in Tab. 2, for case $\beta$=0.2 and *L/d*=4.0, the streamwise lengths of both the wavemaker and SFD base flow recirculation region increase monotonically with Reynolds number, following the same trend as that reported for confined flow past a single cylinder[42]. Besides, the values $\eta$ of the wavemaker near the gap region between the two cylinders are consistently larger than those on the outer sides of the cylinders. Similarly, an increase in the streamwise lengths of the wavemaker and the base flow recirculation region is associated with reduced flow stability.

These results imply that, when the flow configuration changes, the spatial distribution of sensors can be adjusted accordingly to capture the dominant oscillatory information of the flow field more effectively. For example, as *Re* increases, perturbations farther downstream of the two cylinders and closer to the gap region should be monitored to address the shift of the wavemaker.

### 3.2 Effect of blockage ratio

At a fixed spacing ratio of *L/d*=2.0, the blockage ratio is varied by changing the diameter *d* of each cylinder. The Reynolds number is varied by changing the fluid viscosity. The results show that the influence of the blockage ratio $\beta$ on confined flow past side-by-side cylinders is more complex than that on confined flow past a single circular cylinder.

The effects of the blockage ratio $\beta$ on the critical Reynolds number and eigenfrequency are shown in Fig. 11, i.e., phase diagram of the dominant unstable modes. As shown in Fig. 11 (a), the $Re_c$ of both eigenmodes A and B increase progressively with increasing $\beta$, and the rate of increase also becomes larger. At small blockage ratios ($\beta \leq 0.1$), eigenmode B is not excited. For $\beta \leq 0.53$, eigenmode A remains more unstable than eigenmode B and is therefore the dominant eigenmode. At $\beta$=0.53~0.55, the $Re_c$ of eigenmode A exceeds that of eigenmode B, and eigenmode B becomes the dominant unstable mode. As shown in Fig. 11 (*b*), the eigenfrequency decreases gradually with increasing $\beta$, and eventually becomes nearly insensitive to further changes in $\beta$. Although this dependence is nonlinear, it remains monotonic. Therefore, compared with the spacing ratio, the blockage ratio provides a simpler passive control parameter for regulating the flow.

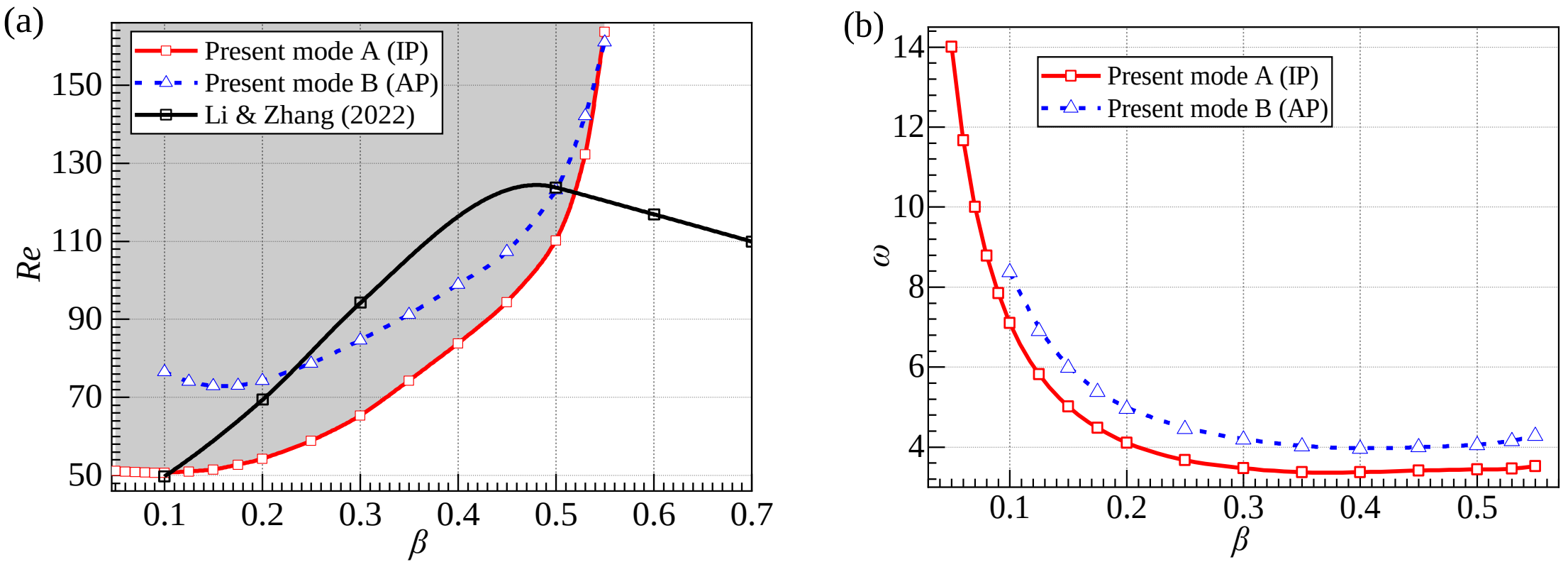

Fig. 11 The effect of blockage ratio $\beta$ on neutral stable curves in the $\beta$-$Re$ (a) and $\beta$-$\omega$ (b) parameter planes at $L/d$=2.0. The grey shaded area in panel (a) denotes unsteady flow patterns undergoing Hopf bifurcation. The confined single-cylinder flow data are from Li & Zhang[42].

### 3.2.1 Influence on global stability modes

Two distinct unstable eigenmodes A and B are identified, as shown in Fig. 12(a, b). Here, the lower-frequency eigenmode A is symmetric in vorticity about axis $y$=0, corresponding to the in-phase (IP) mode, whereas the higher-frequency eigenmode B is antisymmetric and corresponds to the anti-phase (AP) mode. The combined effect of the two unstable eigenmodes gives rise to a multifrequency oscillatory wake, denoted as wake P-V, as shown in Fig. 12(c). From the lift-coefficient histories and the Tot. $c_l$–Tot. $c_d$ phase diagram shown in Fig. 12(d), wake P-V exhibits features characteristic of a chaotic state.

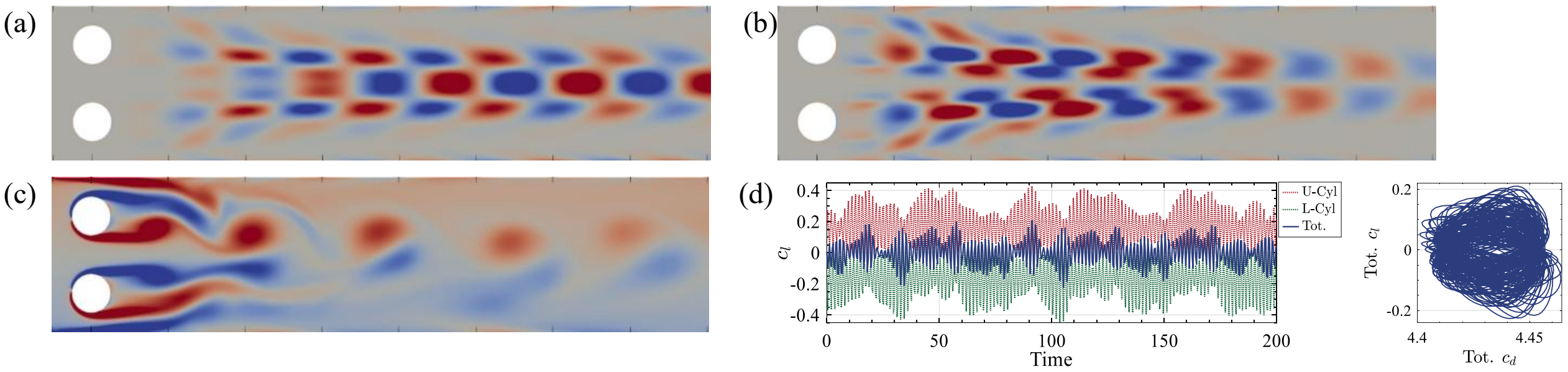

Fig. 12 Linear (a, b) and nonlinear (c, d) results for $\beta$=0.5, $Re$=150, and $L/d$=2. (a) Eigenmode A (IP) ($\sigma$=0.2702, $\omega$=3.4935); (b) eigenmode B (AP) ($\sigma$=0.1784, $\omega$=4.0671); (c) instantaneous vorticity field of chaotic wake P-V; and (d) the lift and drag coefficients of wake pattern V. Where 'Tot.' denotes the sum for the two cylinders, 'U-Cyl' and 'L-Cyl' denote the upper and lower cylinders, respectively.

### 3.2.2 Influence on structural sensitivity

Tab. 3 and Tab. 4 show the wavemaker distributions of eigenmodes A and B in the range of $0.1<\beta<0.5$ and *Re*<180, covering both subcritical and supercritical cases. The curves and contours are visualized in the same manner as in Tab. 1. The wavemaker of eigenmodes A and B overlap strongly in both location and size. Therefore, the instability information associated with both modes can be captured simultaneously using the same sensor-placement strategy.

As shown in Tab. 3, for $L/d$=2.0 and *Re*=75, the streamwise lengths measured in the computational coordinate system of both the wavemaker and the base flow recirculation region first increase and then decrease as $\beta$ increases. It is worth noting, however, that their dimensionless streamwise lengths normalized by the cylinder diameter decrease monotonically with increasing $\beta$. A more general conclusion can therefore be drawn: a reduction in the dimensionless streamwise length is associated with enhanced flow stability. For both eigenmodes A and B, $\eta_{max}$ is consistently located in the gap region between the two cylinders, and this feature becomes more pronounced at larger $\beta$. Similarly, as the $\beta$ increases, the wall-induced deceleration of the fluid on the outer sides of the cylinders becomes stronger, while the relative velocity in the gap region becomes larger, making the gap flow more sensitive to perturbations.

Tab. 3 Effect of $\beta$ on the structural sensitivity at *Re*=75.

| Mode A (Sym) | Mode B (AS) |
| --- | --- |

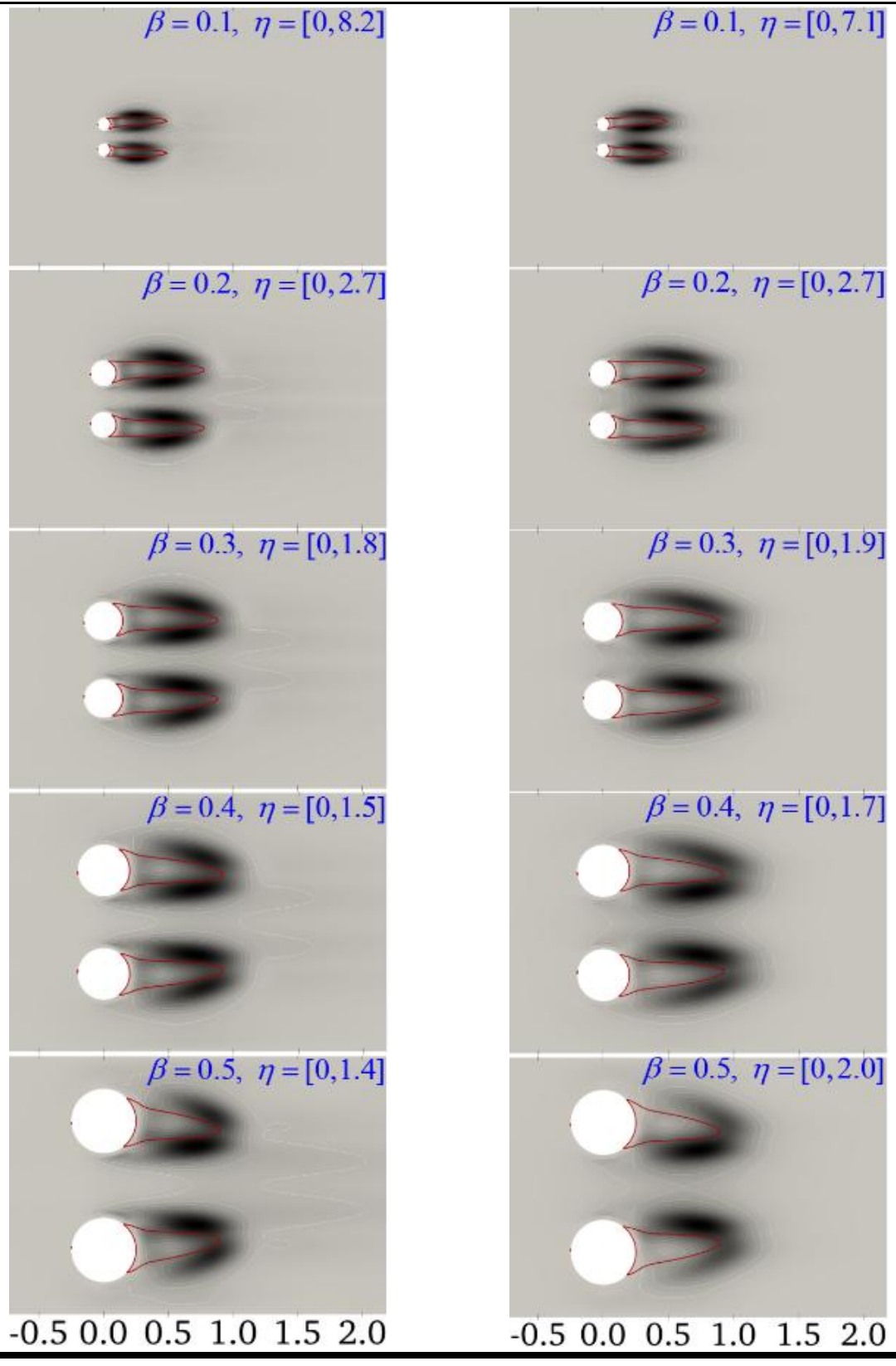

Tab. 4 Effect of Re on the structural sensitivity at $\beta$=0.5.

| Mode A (Sym) | Mode B (AS) |
| --- | --- |

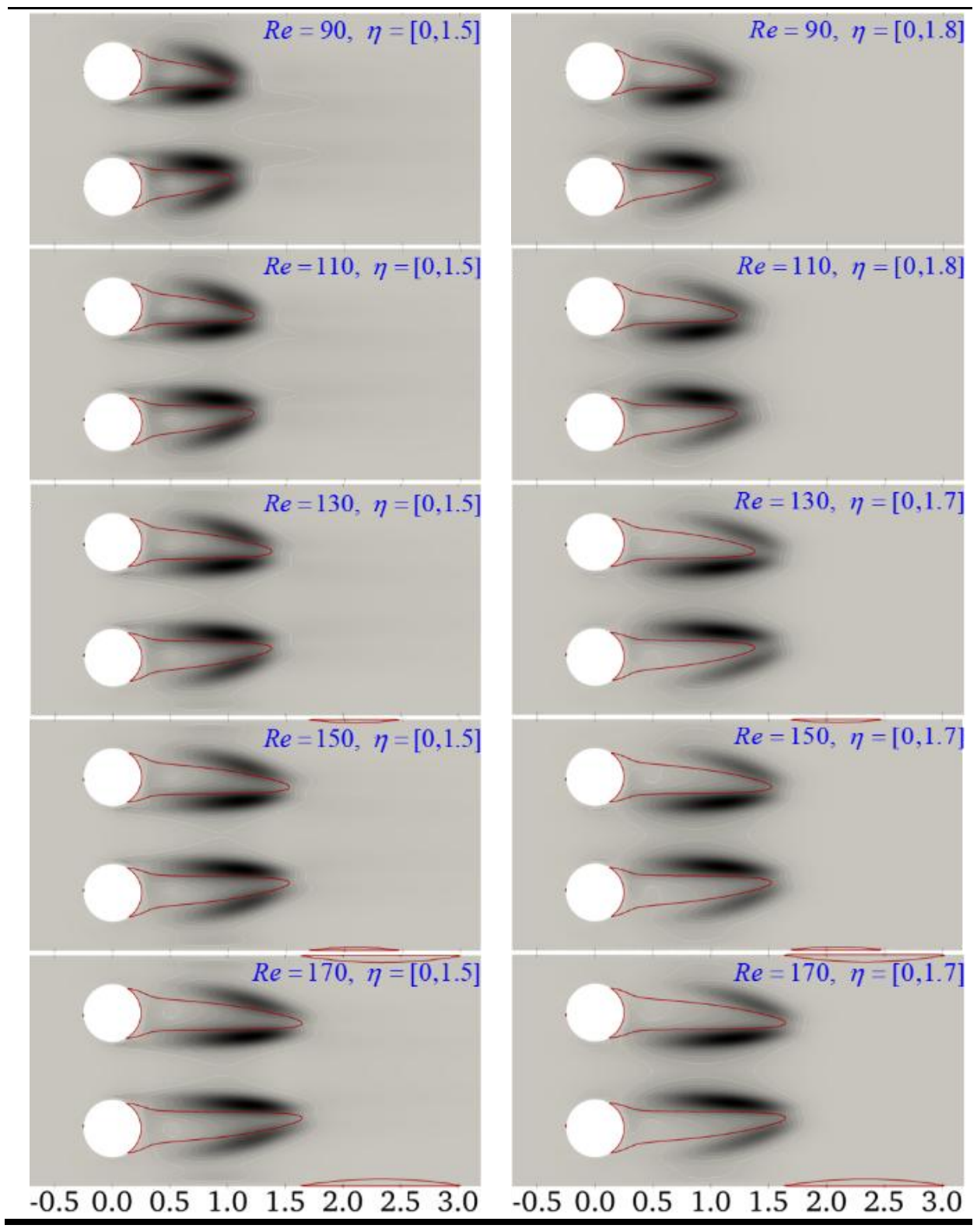

As shown in Tab. 4, for $L/d$=2.0 and $\beta$=0.5, the streamwise lengths of both the wavemaker and the base-flow recirculation region increase with increasing *Re*. At *Re*=150 and 170, recirculation bubbles appear near the channel walls, and this phenomenon has also been observed for confined flow past a single cylinder at large $\beta$[42], where it can stabilize the cylinder wake. However, the second critical Reynolds number for instability associated with this phenomenon in the confined single-cylinder flow is not observed within the parameter space considered in the present study. Similarly, the maximum sensitivity $\eta_{max}$ of eigenmodes A and B is always located in the gap region of the wake between the two cylinders, and shifts downstream as *Re* increases.

## 4. DRL-based AFC results analysis

Complex interactions arise between side-by-side cylinders in a confined domain, and the control strategy must account for feedback from both local and global flow regions. Based on the method validation and the analysis of the flow physics presented above, this section selects supercritical flow states at different spacing ratios after they have evolved into nonlinear saturated vortex-shedding oscillations, and applies AFC to these cases. Different sensor-placement configurations are designed, and multiple control policies are trained and tested to compare their control performance.

### 4.1 Effect of sensor distribution

The sensor placement needs to balance representativeness and computational efficiency. A properly designed sensor distribution can not only capture key flow-field information, but also reduce redundant data and improve the agent's perception of the environment. By using flow-field data sampled from a finite number of sensors as the environmental-state input to the agent, the dimensionality of the data can be substantially reduced while retaining the dominant features of the flow. This provides a feasible set of input data for the agent to learn and optimize different control policies.

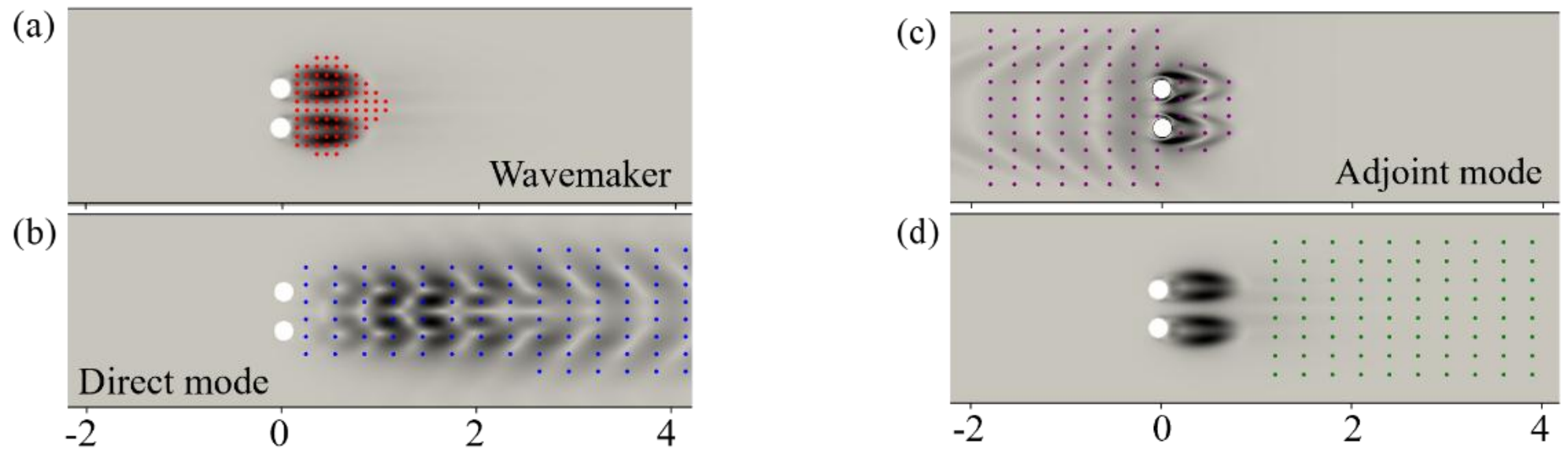

Fig. 13 Four sensor-placement configurations at *L/d*=2.0, *β*=0.2 and *Re*=66. Sensor placement: (a) wavemaker region; (b) direct-mode region; (c) adjoint-mode region; and (d) downstream of the wavemaker.

For the confined side-by-side cylinder flow in the case with *L/d*=2.0, *β*=0.2 and *Re*=66, the wavemaker, direct mode and adjoint mode are shown by the black contours in Fig. 13 (a, b and c), respectively. Four sensor-placement configurations, indicated by the solid dots, are considered: all configurations in Fig. 13(a–d) use 96 sensors. Then, each configuration is trained separately using the enhanced reward function defined in Eq. (12). After 500 training episodes, the time histories of the vortex-shedding energy, drag coefficient and lift coefficient of the confined side-by-side cylinder flow during the testing stage are shown in Fig. 14. It can be observed that the control performance differs significantly among the different sensor configurations.

With sensors configuration (*a*), the vortex-shedding energy decreases most rapidly and falls below 0.03 after $t$>24. This strategy gives the lowest energy level among the four configurations. The sensors configuration (b) gives the poorest control performance, although the direct mode describes the spatial structure of unsteady vortex shedding. The vortex-shedding energy reaches its minimum after $t$>16, but remains approximately 50% of its initial value and exhibits pronounced fluctuations. The control performance of sensors configuration (c) is slightly lower than that of configuration (a). From the sensor distributions shown in Fig. 13 (a) and (c), this may be partly because the sensors covering the adjoint-mode region also include part of the wavemaker region. It is worth noting that sensors configuration (d) fails to produce an effective control policy for suppressing vortex shedding. This indicates that monitoring only the downstream wake region is insufficient for training an effective control strategy.

In summary, by monitoring variations in the core instability region, sensors configuration (a) provides a more effective means of suppressing vortex shedding.

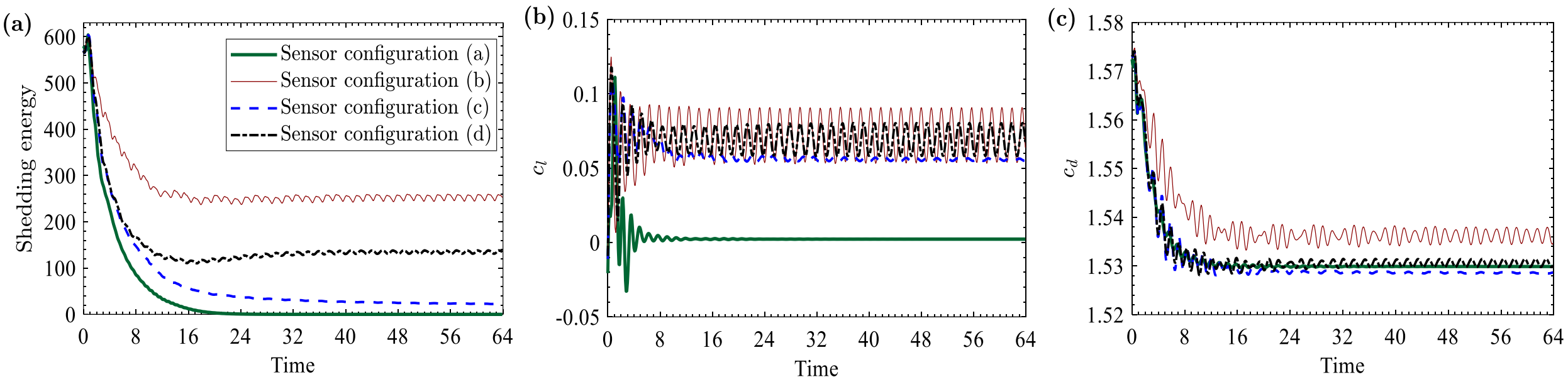

Fig. 14 Time histories of (a) the vortex shedding energy, (b) lift coefficient and (c) drag coefficient during the testing stage for the sensor configurations shown in Fig.13 at the case $L/d$=2.0, $\beta$=0.2 and $Re$=66.

In addition, as shown in Fig. 14 (c), all sensor-placement configurations achieve drag reduction. Under sensors configuration (a), the fluctuation amplitudes of both the lift and drag coefficients are reduced to within ($10^{-5}$), and the time-averaged drag coefficient is reduced by 2.8%. Although sensors configuration (c) does not provide the strongest suppression of lift oscillations, it yields the best drag-reduction performance, with the drag reduced by approximately 2.9%. This indicates that, although suppressing flow oscillations can reduce drag, stronger oscillation suppression does not necessarily lead to a larger drag reduction.

## 4.2 Effect of reward function

Furthermore, this section examines the effects of reward function on control performance. Based on sensors configuration (a) in Fig. 13, as described in the previous section, the time histories of the flow parameters during the training and testing stages are compared for the base reward ($P_e$=800) and enhanced reward functions, as shown in Fig. 15 and Fig. 16.

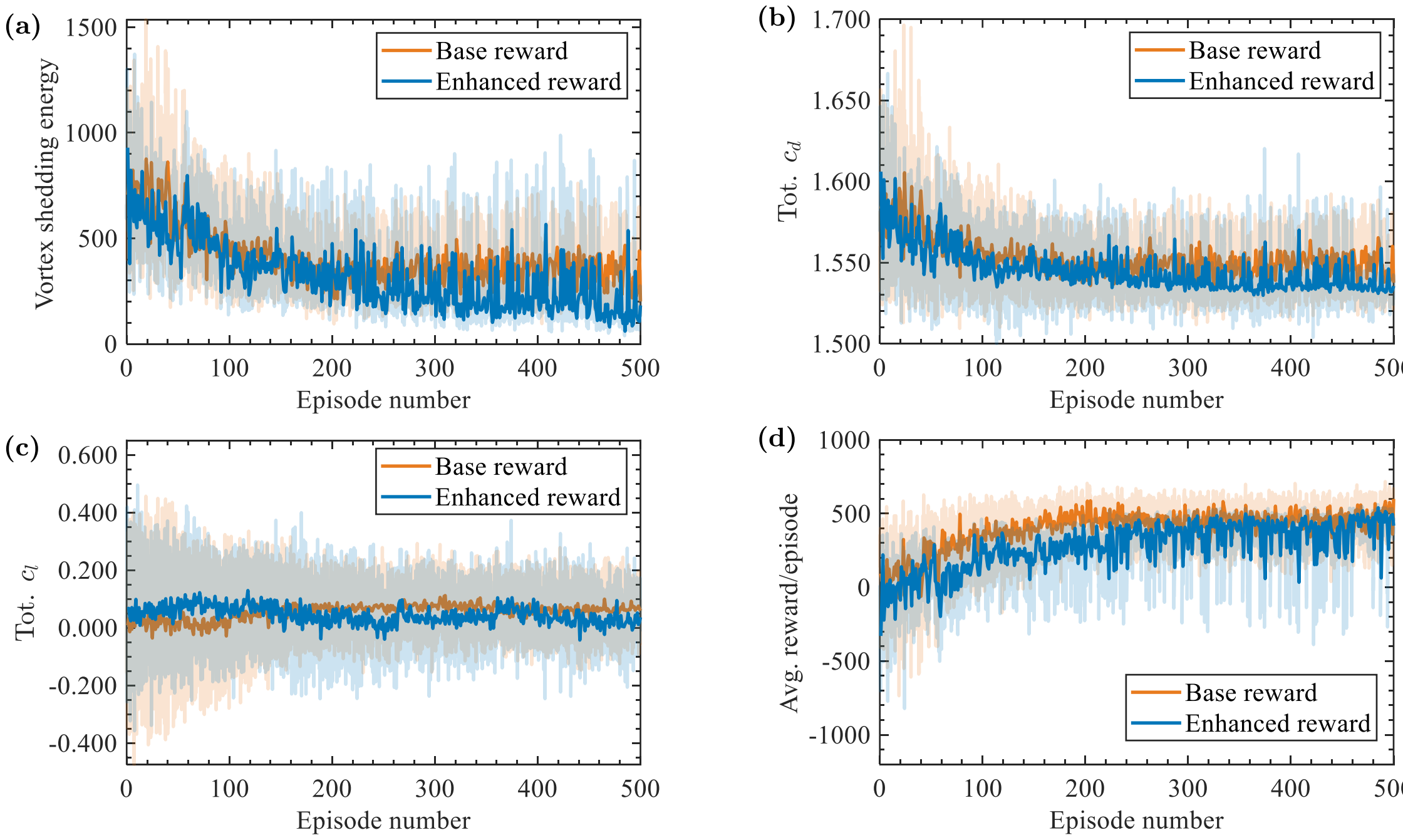

Fig. 15 Preformance comparison of the two reward functions during the training stage. Panel (a): vortex shedding energy, (b): total lift coefficient, (c): total drag coefficient, and (d): time averaged reward/episode. The semi-transparent curves represent the values recorded after each action is executed.

As shown in Fig. 15 (a), before 225 episodes, the vortex-shedding energy obtained using both the base reward and enhanced reward functions gradually decrease with time. After 250 training episodes, the energy obtained using the base reward function no longer decreases in an overall sense, whereas that obtained using the enhanced reward function continues to decrease to a lower level, albeit with larger fluctuation amplitudes. Together with the evolution of the reward shown in Fig. 15 (d), this indicates that, when the base reward function is used, the agent does not sufficiently explore the environment and therefore converges prematurely to a locally optimal policy, resulting in a relatively low converged reward. Fig. 15 (b, c) further show that, when the enhanced reward function is used, the drag coefficient converges to a lower value and the mean value of the post-convergence lift-coefficient fluctuations is also reduced.

Fig. 16 compares the testing results obtained using the two reward functions. With the base reward function, the vortex-shedding energy decreases slowly, and the amplitude of the lift coefficient is gradually reduced; however, the time-averaged value of the lift increases progressively. At the end of the testing stage, the shedding energy is reduced only to 33, indicating that, when the base reward function is used with a penalty value of 800, the agent fails to identify an optimal control policy. By contrast, the enhanced reward function leads to a substantial improvement in the suppression of the vortex-shedding energy. The vortex-shedding energy decreases from its initial value of 563 to zero, and this control objective is achieved as early as $t$=21.

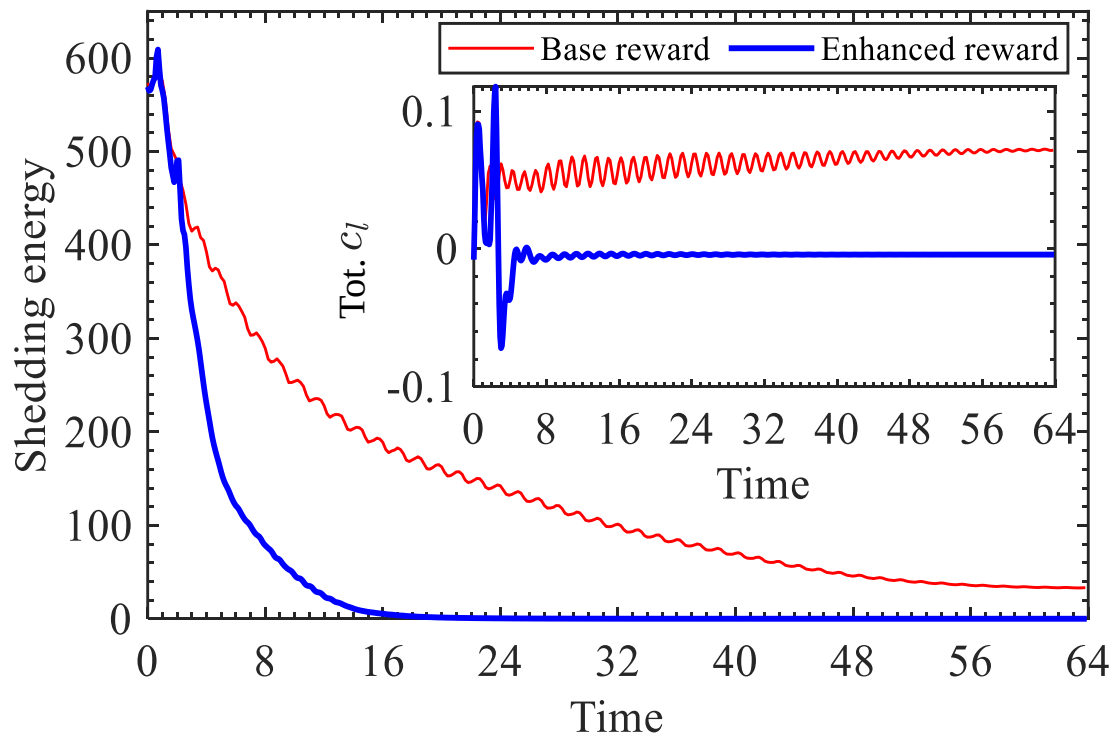

Fig. 16 Comparison of the vortex-shedding energy and total lift coefficient for two reward functions in testing-stage.

### 4.3 Controls on different configurations

This section tests the effectiveness of the proposed AFC strategy on different configurations. The enhanced reward function is used, together with the optimal sensor-placement configuration (a) in Fig. 13 (a). With $\beta$=0.2 and $Re$=66 fixed, four spacing ratios, $L/d$=1.5, 2.0, 2.5 and 3.0, are considered. As shown in Fig. 9 and Fig. 5, these cases correspond to four different flow regimes: (1) a symmetric wake with a linear stable eigenmode; (2) a symmetric wake with a linear unstable eigenmode; (3) an asymmetric wake with two linear unstable eigenmodes; and (4) an intermittently asymmetric wake with two linear unstable eigenmodes, respectively.

For the case ($L/d$=1.5, $\beta$=0.2 and $Re$=66), the flow is stable. Therefore, the nonlinear saturated state of its corresponding supercritical case, ($L/d$=1.5, $\beta$=0.2 and $Re$=99), is used as the initial state for active control. The purpose is to test whether the control strategy over-intervenes in the flow and destabilizes an originally stable state. For the other three configurations $L/d$=2.0, 2.5 and 3.0, their respective nonlinear saturated states are used as the initial states for active control. The flow-control results are shown in Fig. 17 and Fig. 18. To facilitate comparison of vortex-shedding suppression across different flow conditions, the shedding energy time histories shown in Fig. 17 (b) and Fig. 21(b) was normalized by its initial value for each case.

Fig. 18 shows the vorticity contours of the controlled flow at ($t$=64) for the four configurations. It can be seen that, for the cases with spacing ratios $L/d$=1.5 and 2.0, the wake oscillations are effectively suppressed and the flow is driven to a steady state. Correspondingly, both the normalized shedding energy and the lift coefficient are reduced to zero, as shown in Fig. 17. These results indicate that the present control strategy can effectively handle both stable and unstable flow configurations, without destabilizing flows that are originally stable.

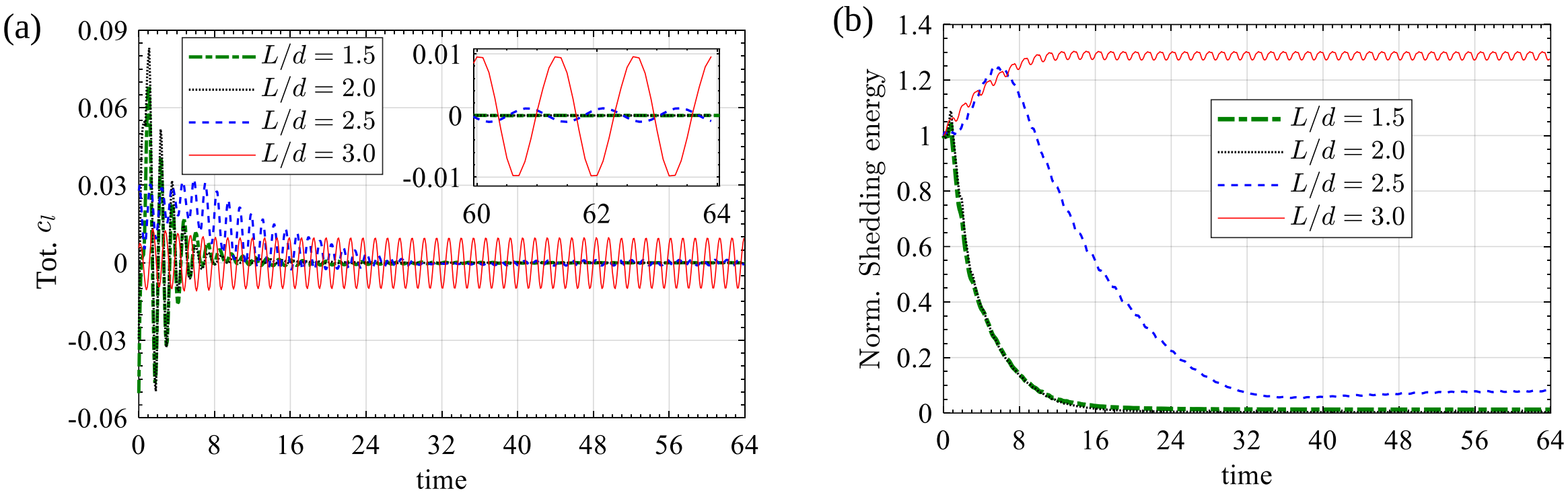

Fig. 17 Testing-stage control results for confined side-by-side cylinder flows at different spacing ratios at *β*=0.2 and *Re*=66 with single action. Time histories of (a) the total lift coefficient and (b) the normalized vortex-shedding energy.

As shown in Fig. 18 (c), slightly periodic oscillations remain in the flow field for the *L/d*=2.5 configuration, and a slight asymmetry is still present in the wakes. Fig. 17 further shows that the time histories of the lift coefficient and normalized shedding energy still exhibit weak fluctuations, indicating that the oscillations are not completely suppressed. Nevertheless, the vortex-shedding energy is substantially reduced, by approximately 91%. A similar difficulty has also been reported for DRL-based AFC under weakly turbulent conditions[35] with intermittency, where complete suppression of vortex shedding was also difficult to achieve. Nevertheless, the agent was able to reduce drag by up to 30%.

Fig. 18 (d) shows the controlled wake state for the configuration with *L/d*=3.0. It can be seen that pronounced spatial oscillations still persist in the controlled flow field. Moreover, Fig. 17 (b) shows that the controlled vortex-shedding energy is even higher than that of the uncontrolled flow, indicating that present AFC method has a detrimental effect in this case. The wake for this configuration exhibits strong asymmetry (see Fig. 9 c), and the symmetric and symmetry-broken states compete intermittently for dominance in the flow, which corresponds to a strong bistability[11]. In the next section, we refine the control strategy based on physical insights from the structural sensitivity analysis to control the wake at moderate spacing ratios (*L/d*=2.5, 3.0), where intermittent symmetry breaking and flip-flopping behaviour occur.

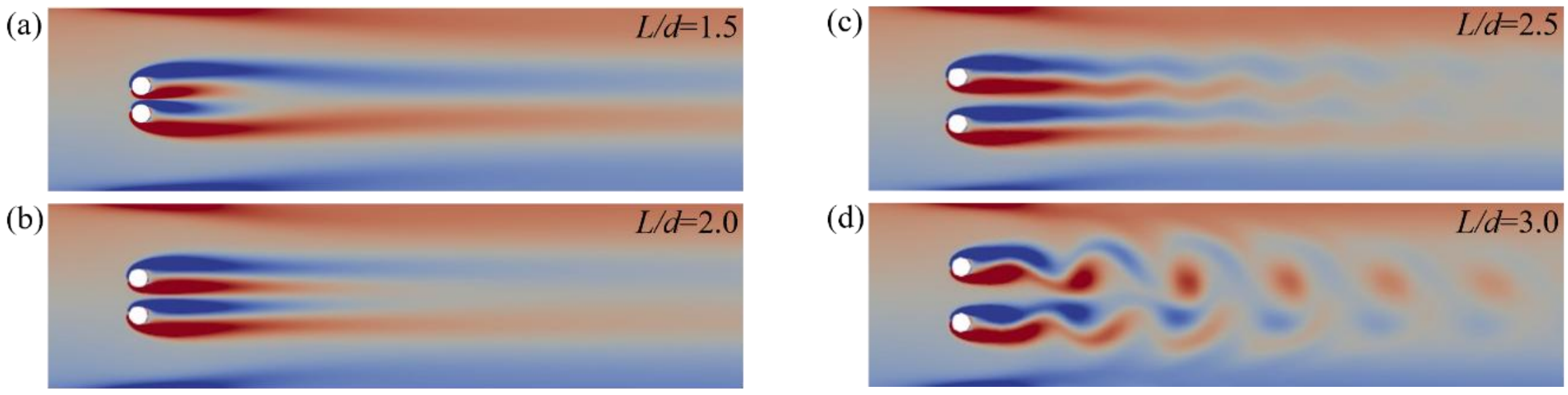

Fig. 18 Vorticity contours snapshots of the controlled flow at *t*=64 for the case *β*=0.2 and *Re*=66.

## 4.4 Two-action control of bistable and chaos flow

To address the inability of the previous control strategy to restore flow symmetry in the bistable regimes at moderate spacing ratios (*L/d*=2.5 and 3.0, *β*=0.2) and the chaotic wake P-V at high blockage ratios (shown in Fig. 12 c), this section introduces a stability-guided improvement to the control strategy, drawing also on multi-action control approaches developed for complex chaotic flows[60]. The structural sensitivity results in Tab. 1 show that, as *L/d* increases, the location of the maximum wavemaker sensitivity, $\eta_{max}$, shifts from the outer sides of the cylinder pair towards the gap region, suggesting that the recirculation region downstream of the gap plays an increasingly important role in flow stability. Guided by this observation, two additional synthetic-jet orifices are introduced in the gap region to enable two independent control actions, as shown in Fig. 19 (a). The blowing and suction through jets 1 and 2 on the upper cylinder define action 1, whereas those through jets 3 and 4 on the lower cylinder define action 2. Each action satisfies the zero-net-mass-flux constraint.

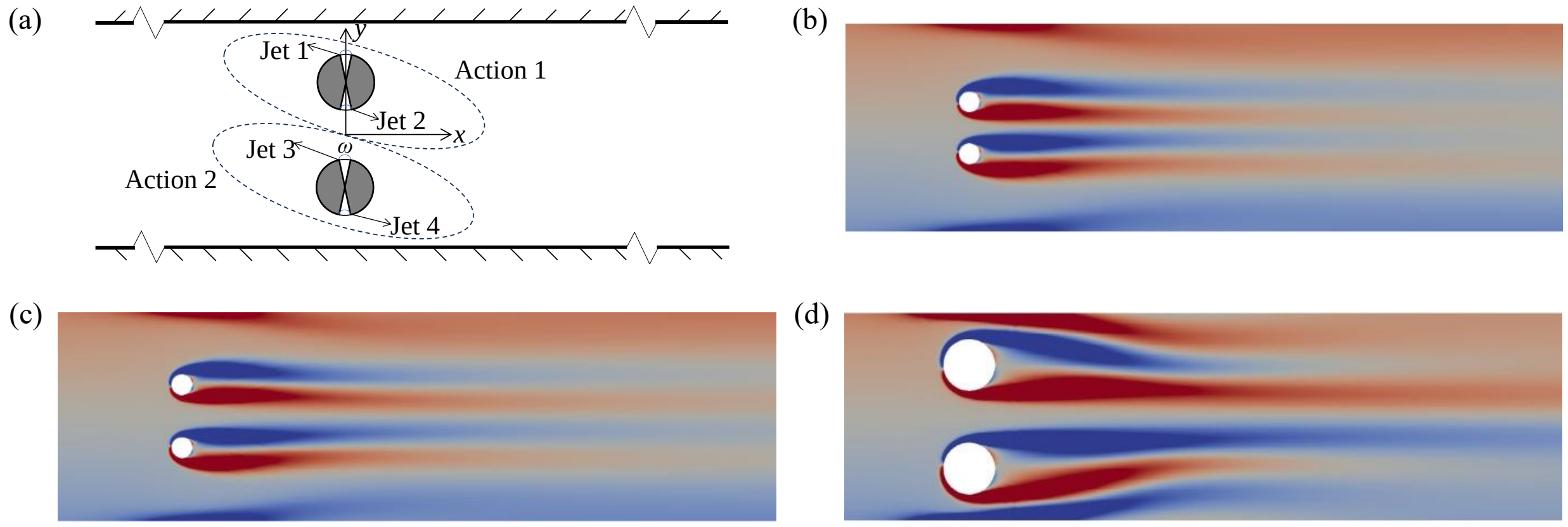

Fig. 19 (a) Configuration of the two control actions. Vorticity fields during the testing stage at *t*=64, showing nearly complete suppression of wake oscillations and restoration of flow symmetry for (b) $\beta$=0.2, *Re*=66, *L/d*=2.5, (c) $\beta$=0.2, *Re*=75, *L/d*=3.0 and (d) $\beta$=0.5, *Re*=150, *L/d*=2.

For *L/d*=2.5, Fig. 20 (a–d) compare the vortex-shedding energy, drag coefficient, lift coefficient and reward, respectively, for the one- and two-action control schemes. Over the first 400 episodes, the two-action scheme achieves lower vortex-shedding energy and a correspondingly lower drag coefficient than the single-action scheme, consistent with physical causality. This suggests that the single-action agent initially explores less effective policies and gradually identifies an effective shedding-suppression strategy through subsequent gradient-based policy updates. Fig. 21 (c) indicates that the two-action agent learns cylinder-specific actuation strategies: a larger jet flow-rate magnitude is used to control the larger vortices behind the upper cylinder, and a smaller magnitude to control the smaller vortices behind the lower cylinder in Fig. 18 (c). This targeted actuation enables the two-action scheme to outperform the single-action scheme from the early stages of training in Fig. 20(a). During testing, the two-action scheme reduces the vortex-shedding energy to zero (see Fig. 20a), overcoming the failure of the single-action strategy. The final controled vorticity field at *t*=64 in Fig. 19 (b) further demonstrates more complete suppression of wake oscillations than that achieved with a single action in Fig. 18 (c).

The two-action scheme is further applied to the *L/d*=3.0 configuration, using a more unstable case with a higher oscillation frequency (*Re*=75, $\sigma$=0.2885, $\omega$=0.7576). The initial flow state for control resembles that shown in Fig. 9(c), with the wake behind the side-by-side cylinders switching intermittently between the asymmetric pattern P-III and the symmetric pattern P-IV. Unlike the unsuccessful single-action control shown in Fig. 17, the two-action scheme completely suppresses the oscillations in this more unstable flow, despite its higher initial vortex-shedding energy (Fig. 19c). As in the *L/d*=2.5 case, the jet-flow-rate histories in Fig. 21(c) show that the agent learns two distinct, cylinder-specific actuation strategies. As shown in Fig. 21(b), the normalized vortex-shedding energy approaches zero at *t*≈18 for *L/d*=3.0, compared with *t*≈40 for *L/d*=2.5. The difference in control performance can be understood in terms of the flow complexity associated with wake coupling. As *L/d* increases from 2.0 to 6.0, the coupling between the two cylinder wakes progressively weakens, and the wakes develop increasingly distinct, individual structures. This reduction in wake interaction and overall flow complexity is why the two-action scheme suppresses wake oscillations more rapidly at $L/d$=3.0 than at $L/d$=2.5.

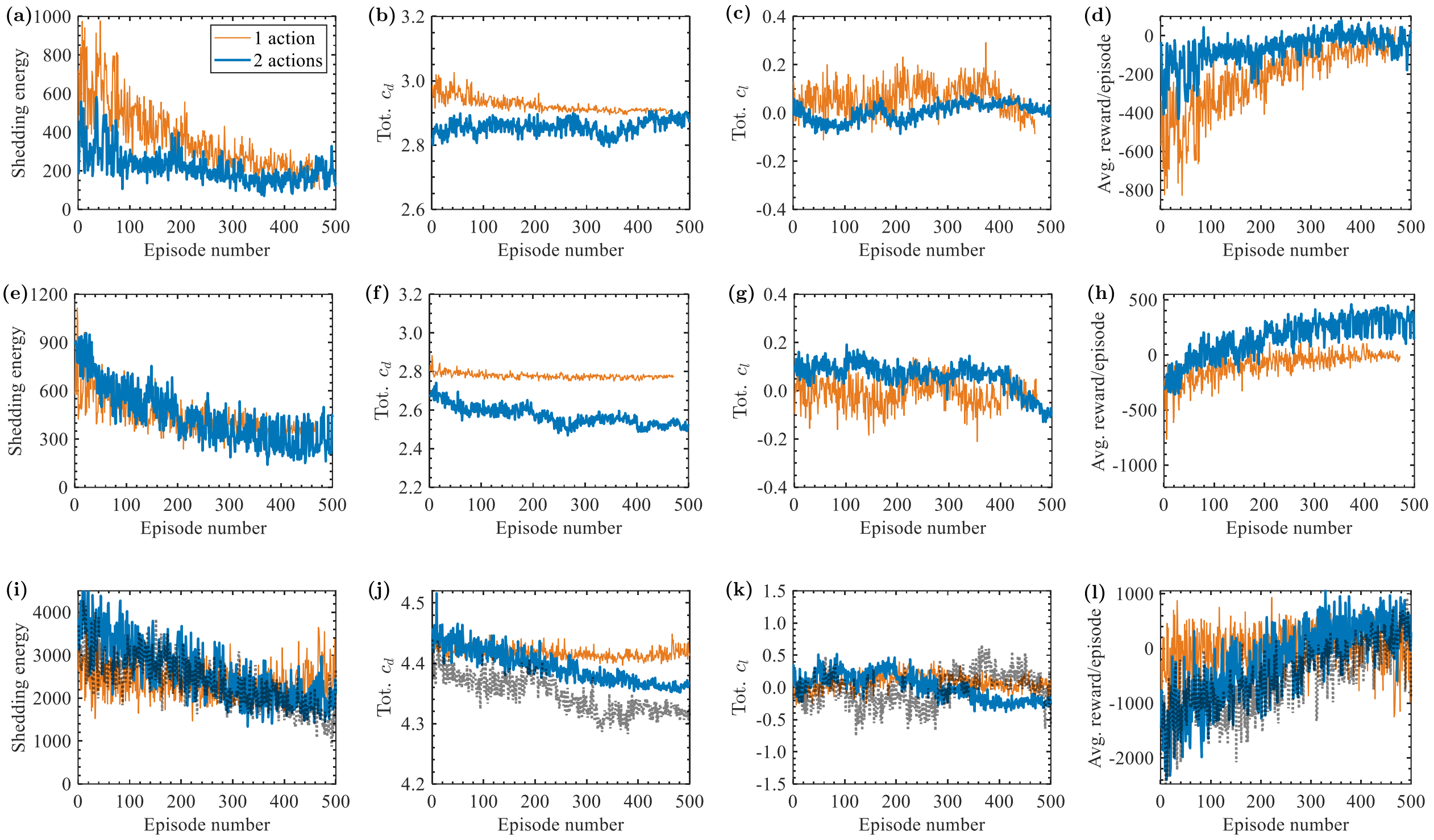

Fig. 20 Effect of the number of control actions on training performance, where the orange and blue solid lines represent single action and two actions, respectively, with the jets located at ±90° as shown in the Fig. 19 (a). Panels (a-d) are of case $\beta$=0.2, $Re$=66, $L/d$=2.5. Panels (e-h) are of case $\beta$=0.2, $Re$=75, $L/d$=3.0. Panels (i-l) are of case $\beta$=0.5, $Re$=150, $L/d$=2, where the semi-transparent black dotted lines represent the two actions case, but the jets are located 9° upstream of the separation points.

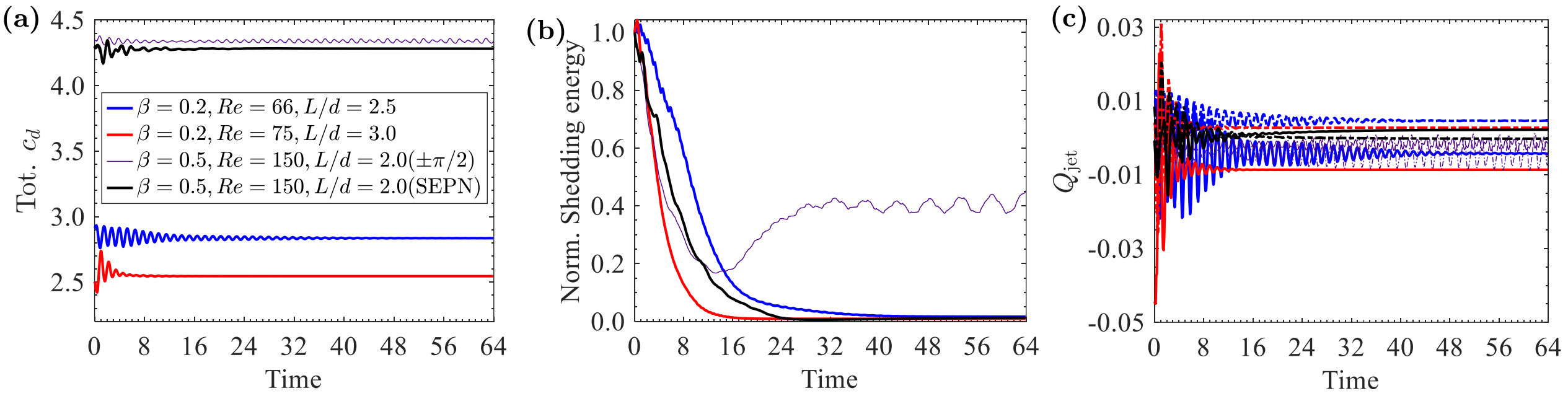

Fig. 21 Testing-stage time histories under two-action control: (a) total drag coefficient Tot. $c_d$, (b) normalized vortex-sheding energy and (c) synthetic-jet flow rate $Q_{\text{jet}}$, with solid and dashed lines representing action 1 and action 2, respectively. In the legend, '±π/2' denotes jets located at ±90° (see Fig. 19 a), and 'SEPN' denotes jets located 9° upstream of the separation points.

For the chaotic wake at $\beta$=0.5, $Re$=150 and $L/d$=2 (Fig. 12), neither the single-action (Fig. 1) nor the two-action configuration (Fig. 19 a), with jets located at ±π/2, effectively suppressed the vortex-shedding energy. The single-action strategy failed to reduce the shedding energy throughout training. Although the two-action strategy learned to reduce the shedding energy during training (Fig. 20 i–l), it achieved only partial suppression during testing, with persistent wake oscillations (Fig. 21 b). Moreover, its performance was highly sensitive to the initial flow field, with the learned policy tending to overfit certain initial conditions and even increasing the shedding energy under others. To overcome these limitations, we drew on physical insights from the resolvent analyses of R. Ma et al. [46-47], who identified an optimal jet location approximately 9° upstream of the separation points. Although their studies employed steady jets, this finding suggests that placing jets upstream of the separation points may enable earlier and more effective intervention in vortex formation and subsequent wake evolution. Guided by this insight, we relocated the synthetic jets to approximately 9° upstream of the separation points while keeping all other settings identical to those of the original two-action configuration at $\pm\pi/2$. The corresponding training and testing results are shown in Fig. 20 (i–l) and Fig. 21, respectively. Compared with the original configuration, the modified strategy achieved more effective suppression of vortex-shedding energy and greater drag reduction during both stages. Notably, it successfully suppressed the chaotic wake oscillations and drove the flow towards the unstable steady base state.

Overall, these results indicate that two independent, cylinder-specific control actions targeting both the outer and gap-side flows are key to suppressing flip-flopping and even chaotic wake dynamics.

## 5. Conclusions

In this study, a research framework combining spectral-element DNS, global linear stability analysis, structural sensitivity analysis and DRL is developed to investigate the instability and AFC of the wake of two side-by-side cylinders in a channel for $Re<180$, $\beta<0.6$ and $1<L/d<6$. Unlike control strategies that rely solely on force signals or flow-field snapshots, the present approach first identifies dominant modes and wavemaker regions responsible for perturbation growth in the confined wake. This information is then utilized in the sensor placement and reward-function design, thereby embedding flow-stability priors into the data-driven control process. The main conclusions are as follows.

At a fixed blockage ratio of $\beta=0.2$, the spacing ratio governs the competition among the global instability modes. For $L/d\leq1.5$, the wake is dominated by a low-frequency, single-bluff-body-type eigenmode. As $L/d$ increases, local cylinder-wake dynamics become increasingly important and the in-phase and anti-phase eigenmodes compete. As $L/d>\sim2.75$, the anti-phase eigenmode becomes dominant. In the range where multiple eigenmodes are simultaneously unstable, the nonlinear wake can switch intermittently between symmetric and symmetry-broken states. At a fixed spacing ratio of $L/d=2.0$, increasing $\beta$ raises the critical $Re$ of both the in-phase and anti-phase eigenmodes and reduces their critical oscillation frequencies; the anti-phase eigenmode becomes dominant only at sufficiently large $\beta$.

Structural sensitivity analysis shows that the wavemakers of the leading modes are concentrated in the near wakes and overlap strongly with the base-flow recirculation regions. A shorter dimensionless streamwise length of the wavemaker and recirculation region is generally associated with greater stability, whereas increasing $Re$ extends these regions downstream and destabilizes the flow. These observations provide a physics-based guideline for sensor placement.

Using sensors distributed over the wavemaker region produced the most effective DRL control. The dynamic reward (referred to as the enhanced reward function in present work), improved convergence, reduced sensitivity to the empirical penalty parameter, and achieved superior control performance compared with the basic reward function. For wakes dominated solely by a Hopf bifurcation, the single-action control policy drove the flow close to the steady base state, nearly eliminated vortex-shedding energy and force fluctuations, and reduced the mean drag by approximately 2.8%. However, the single-action strategy was insufficient to fully control the more complex oscillatory wakes at $L/d=2.5$ and 3.0, characterized by bistability-induced intermittent flip-flopping and symmetry breaking. Guided by qualitative insights from the stability analysis, a two-action control strategy was developed to target the flow more effectively. This strategy completely suppressed vortex shedding while restoring flow symmetry, achieving the same control objectives even for the more complex chaotic wake pattern P-V at high values of both $\beta$ and $Re$. To the best of the authors' knowledge, this represents the first successful demonstration of controlling both bistability-induced wake dynamics and chaotic wakes behind side-by-side cylinders.

**Acknowledgments**

This work was supported by the National Natural Science Foundation of China (No. 12202200 and 12402296) and the Fundamental Research Funds for the Central Universities (No. 30925010501).

**Appendix A: Numerical implementation and verification**

In the study of nonlinear wake dynamics and AFC, DNS are performed using a high-order spectral-element method (SEM), which combines the geometrical flexibility of the finite-element method with the high accuracy of spectral methods. The spatial discretization is based on quadrilateral elements and $N^{\text{th}}$-order Legendre polynomial basis functions defined on Gauss–Lobatto–Legendre (GLL) nodes. The nonlinear terms are treated using an explicit extrapolation scheme, namely the operator-integration-factor splitting (OIFS) method, to maintain computational efficiency, while the viscous terms are advanced implicitly using a second-order backward differentiation formula (BDF2) to improve numerical stability. The time step is chosen such that the CFL condition (CFL≤1) is satisfied, ensuring numerical stability at high spatial resolution, especially in strongly nonlinear regions induced by the local jets, and preventing numerical divergence. In addition, MPI-based parallel computing is employed to accelerate the simulations, which substantially reduces the training time required for reinforcement learning.

The geometry and mesh of the computational domain used in the present simulations are shown in Fig. A1. Following the domain configuration adopted by Li & Zhang[42] for confined flow past a single circular cylinder, the inlet is placed at ($x_1$=5$d$) upstream of the side-by-side cylinders. The outlet is located at ($x_3$=10$d$) downstream of the cylinders, which is sufficiently far to allow the wake to develop. The computational mesh consists mainly of three parts. The first part comprises two rectangular subdomains with streamwise lengths $x_1$ and $x_3$, containing ($n_1$+$n_3$)×2$n_y$ nodes in total. The second part is a square subdomain of streamwise length $x_2$ enclosing the side-by-side cylinders, containing (2$n_2$+4$n_y$)×$n_b$ nodes. The third part is an O-type refined mesh region around each cylinder, as shown in Fig. A1 (b), containing 2($n_2$+$n_y$)×$n_c$ nodes.

(a)

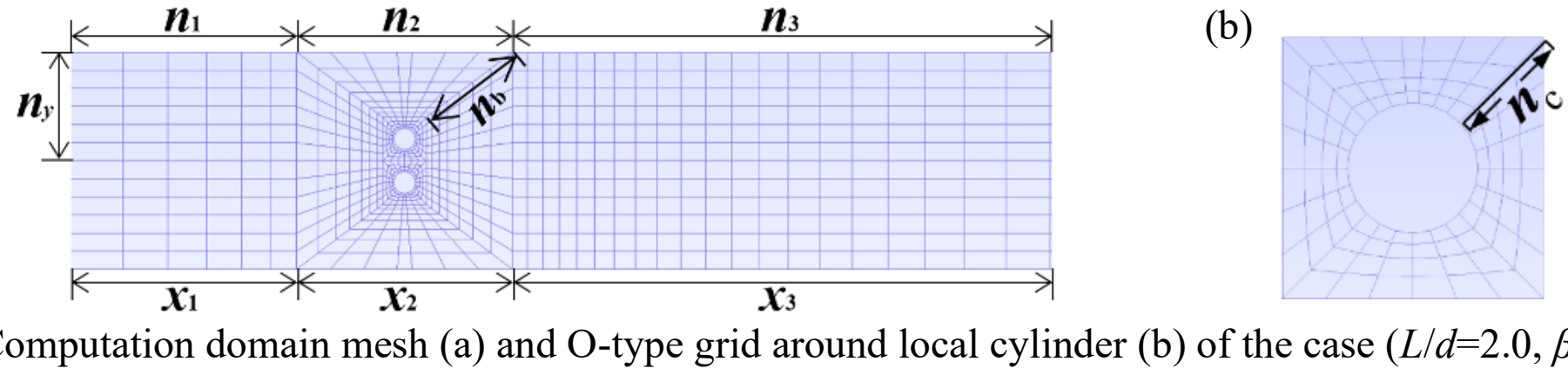

Fig. A1. Computation domain mesh (a) and O-type grid around local cylinder (b) of the case ($L/d$=2.0, $\beta$=0.2).

For the supercritical condition case of $L/d$=2.0, $\beta$=0.2 and $Re$=66, five mesh configurations were designed to assess grid independence, with the spatial resolution increasing progressively from M1 to M5. The time-averaged drag coefficient Tot. $\overline{c}_d$ , the amplitude of the lift coefficient $A$(Tot. $c_l$), and the $St$ based on the lift are reported in Tab. A1. Where $N_{\text{tot.}}$ denotes the total number of elements. For Tot. $\overline{c}_d$ , the largest discrepancy occurs between the coarsest mesh M1 and the finest mesh M5, but is only 0.012%; the discrepancy between M2 and M5 is further reduced to 0.001%. For $A$(Tot. $c_l$), the largest difference is found between M2 and M5, but is only 0.03%, while the discrepancies between the other mesh configurations and M5 are all within 0.005%. For *St*, the largest discrepancy occurs between M2 and M5, reaching 0.85%, whereas the discrepancies between the other configurations and M5 are all within 0.007%. Overall, the number of elements has only a minor influence on Tot. $\overline{c}_d$ and *A*(Tot. $c_l$). Therefore, considering the influence of the element number on *St*, mesh M3 is selected for all simulations in this study.

Tab. A1 Grid independence test ($L/d$=2.0, $\beta$=0.2, *Re*=66)

| No. | $N_{tot.}$ | Tot. $\overline{c}_d$ | *A*(Tot. $c_l$) | *St* by Tot. $c_l$ |
|---|---|---|---|---|
| M1 | 764 | 1.58789 | 0.09942 | 0.14202 |
| M2 | 1044 | 1.58806 | 0.09945 | 0.14204 |
| M3 | 1448 | 1.58806 | 0.09942 | 0.14202 |
| M4 | 2054 | 1.58807 | 0.09942 | 0.14201 |
| M5 | 3108 | 1.58808 | 0.09941 | 0.14202 |

**Appendix B: Validation of the present IRAM and sensitivity analysis codes**

Owing to the lack of available stability-analysis results for confined flow past side-by-side cylinders, the present IRAM code is validated against previous results for the single cylinder flow in a confined domain[42, 61-62]. The $Re_c$ obtained by the present IRAM code for different $\beta$, are listed in Tab. B1. The maximum discrepancy between the present results and those of Li & Zhang[42] is less than 0.7%. Consistently, $Re_c$ reaches its maximum at $\beta$=0.5 and then decreases as $\beta$ is further increased.

Fig. B1 shows the wavemaker regions of the most unstable direct and adjoint eigenmode for the cases at different Reynolds numbers and blockage ratios, along with base flow recirculation region. The comparison shows that their structures are consistent with the these reported in Figure 10 of Li and Zhang[42].

Overall, the comparison demonstrates that the stability results obtained by the present IRAM and sensitivity analysis codes agree very well with previous results[42, 61-62], both quantitatively and in terms of the flow structure.

Tab. B1. Validation using the $Re_c$ of confined flow past a single-cylinder

| $\beta$ | 0.1 | 0.2 | 0.3 | 0.5 | 0.7 |
|---|---|---|---|---|---|
| $Re_c$ (Pre.) | 51.35 | 69.78 | 94.33 | 124.78 | 110.74 |
| $Re_c$[42] | 51.12 | 69.93 | 94.13 | 124.83 | 109.93 |
| Err. | 0.4% | 0.3% | 0.2% | 0.1% | 0.7% |

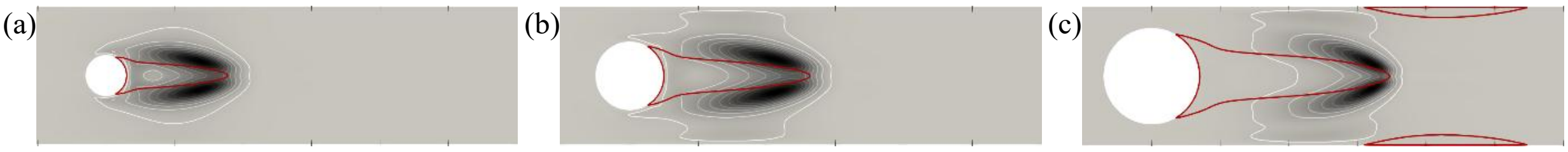

Fig. B1. Wavemaker regions (grey colour) and base flow recirculation region (red solid line) for confined flow past a single cylinder at (a) $\beta$=0.3, *Re*=115; (b) $\beta$=0.5, *Re*=150; and (c) *Re*=185, $\beta$=0.7.